\documentclass{LMCS}

\def\dOi{10(2:8)2014}
\lmcsheading%
{\dOi}
{1--28}
{}
{}
{Dec.~10, 2013}
{Jun.~12, 2014}
{}

\ACMCCS{[{\bf Theory of computation}]: Continuous
  mathematics---Topology---Geometric topology}

\usepackage{amssymb}

\usepackage{hyperref}
\usepackage{graphicx}

\theoremstyle{plain}

\newcommand{\diam}{\mathop{\mathrm{diam}}}

\newcommand{\mesh}{\mathop{\mathrm{mesh}}}
\newcommand{\fdiam}{\mathop{\mathrm{fdiam}}}
\newcommand{\fmesh}{\mathop{\mathrm{fmesh}}}

\begin{document}

\title[Computability of 1-manifolds]{Computability of 1-manifolds}

\author[Konrad Burnik]{Konrad Burnik}   
\address{University of Zagreb, Croatia} 
\email{kburnik@gmail.com, zilj@math.hr}
\author[Zvonko Iljazovi\'{c}]{Zvonko Iljazovi\'{c}}   
\address{\vspace{-18 pt}} 




\begin{abstract}
  \noindent    A semi-computable set $S$ in a computable metric space
  need not be computable. However, in some cases, if $S$ has
  certain
  topological properties, we can conclude that $S$ is computable.
  It is known that if a semi-computable set $S$ is a compact manifold with boundary, then
  the computability of $\partial S$  implies the
  computability of $S$. In this paper we examine the case when $S$ is a 1-manifold with
  boundary, not necessarily compact. We show that a similar result holds in this case
  under assumption that $S$ has finitely many components.
\end{abstract}

\maketitle

\section{Introduction}\label{introd}
A closed subset of $\mathbb{R}^{m} $ is computable if it can be
effectively approximated by a finite set of points with rational
coordinates with arbitrary precision on an arbitrary  bounded
region of $\mathbb{R}^{m} $.
 A compact subset $S$ of
$\mathbb{R}^{m} $ is semi-computable if we can effectively enumerate
all rational open sets which cover $S$.  Each
compact computable set is semi-computable. On the other hand, there exist semi-computable
sets  which are not computable.

Hence the implication
\begin{equation}\label{intro-2}
S\mbox{ semi-computable } \Rightarrow S \mbox{ computable
}
\end{equation}
does not hold in general and the question arises whether there are
some conditions under which it does hold. A motivation for this
question lies in the fact that semi computable subsets of
$\mathbb{R}^{m} $ are exactly compact co-computably enumerable
sets. A closed subset of $\mathbb{R}^{m} $ is called co-computably
enumerable (co-c.e.)\ if its complement can be effectively covered
by open balls. Furthermore, co-c.e.\ sets are exactly the sets of
the form $f^{-1} (\{0\})$, where $f:\mathbb{R}^{m}\rightarrow
\mathbb{R}$ is a computable function. So the question under what
conditions (\ref{intro-2}) holds is related to the question under
what conditions the set of all zero-points of a computable
function $f:\mathbb{R}^{m}\rightarrow \mathbb{R}$ is computable.

 It is known that there exists a computable function
$f:\mathbb{R}\rightarrow \mathbb{R}$ which has zero-points and all of them lie in $[0,1]$, but
none of them is computable \cite{sp:mib}. This means that $f^{-1} (\{0\})$ is a nonempty semi-computable set
which contains no computable point. In particular, $f^{-1} (\{0\})$ is not computable.
Since each nonempty
computable set contains computable points, this shows that there exist semi-computable sets
which are ``far away from being computable''.

However, it turns out that under certain assumptions  implication
(\ref{intro-2}) does hold. In particular, it has been proved in
\cite{mi:mib} that (\ref{intro-2}) holds
 whenever $S\subseteq \mathbb{R}^{m} $ is a topological sphere (i.e.\ homeomorphic to
 the unit sphere $S^{n}\subseteq \mathbb{R}^{n+1}$ for some $n$) or $S$ is homeomorphic to the closed unit ball $B^{n}\subseteq \mathbb{R}^{n}$ for some $n$
(i.e.\ $S$ is an $n-$cell) by a homeomorphism $f:B^{n} \rightarrow
S$ such that $f(S^{n-1})$
 is a semi-computable set.
Furthermore, by \cite{lmcs:mib}, these results hold not just in
$\mathbb{R}^{m}$, but  also in any
 computable metric space which is locally computable.
  Results related to (\ref{intro-2}) can also be found
 in
 \cite{br:mib}, \cite{zi:mib}, \cite{kihara} and \cite{ziegler}.

Recently, the results for
 topological spheres and cells with semi-computable boundary spheres
have been generalized in \cite{lmcs:mnf} where it was proved that
in any computable metric space implication (\ref{intro-2}) holds
if $S$ is a compact manifold with boundary such that the boundary
$\partial S$ is computable. In other words, if $S$ is a compact
manifold with boundary and if $S$ is semi-computable, then
\begin{equation}\label{intro-3}
\partial S\mbox{ computable } \Rightarrow S \mbox{ computable.
}
\end{equation}
The notion of a semi-computable set coincides with the notion of a
compact co-c.e.\ set in a computable metric space which has
compact closed balls and the effective covering property.
Therefore, in such a computable metric space, if $S$ is a compact
manifold with boundary and if $S$ is co-c.e., then (\ref{intro-3})
holds.

In this paper we observe the case when $S$ is a 1-manifold, not
necessarily compact, and we examine what can be said in this case
in view of implication (\ref{intro-3}). We first have to find some
appropriate generalization of the notion of a semi-computable
compact set. The idea is that this new notion be a generalization
to those sets $S$ which may not be compact, but such that $S\cap
B$ is compact for each closed ball $B$.
 We will say that $S$ is semi-computable compact on closed balls or semi-c.c.b.\ if $S\cap B$
is semi-computable, uniformly for each closed rational ball $B$ in the ambient space.

Our main result will be this: if $S$ is a 1-manifold with boundary
in a computable metric space and if $S$ is semi-c.c.b.\ and $S$
has finitely many components, then (\ref{intro-3}) holds. We will
also show that (\ref{intro-3}) does not hold in general  (without
the assumption that $S$ has finitely many components).

 It will turn out that
in a computable metric space which has compact closed balls and the effective covering property the notions
of a semi-c.c.b.\ set and a co-c.e.\ set coincide. Therefore, in such a computable metric space
 we will have
that if $S$ is a  1-manifold
with boundary, $S$ is co-c.e.\ and $S$ has finitely many components, ž
then (\ref{intro-3}) holds.

The main step in the proof of our main result is to prove the following:
if $S$ is homeomorphic to $[0,\infty\rangle $ by a homeomorphism which
 maps $0$ to a computable point or $S$ is homeomorphic to
 $\mathbb{R}$, then $S$ is computable if it is semi-c.c.b. (Here $[0,\infty\rangle$
 denotes the set of all nonnegative real numbers.) Moreover,
 we will prove the following: if $S$ is such a set and $S\cup F$ is semi-c.c.b.,
  where $F$ is closed and disjoint
 with $S$, then $S$ is a computably enumerable set, which means that we can effectively enumerate all
 open rational balls which intersect $S$. This will be the key result and it will
 easily imply the main result for 1-manifolds.

In order to prove this, the central notion will be the notion of a
chain and we will rely on techniques from \cite{zi:mib}.

It should be mentioned here that a semi-c.c.b.\ 1-manifold with boundary (with finitely many components)
 need not be computable if its boundary
is not computable.
An example for this we already have in the compact case: in each $\mathbb{R}^{m} $ there exists a line segment
which is semi-computable, but not computable \cite{mi:mib} (of course, at least one endpoint of such a line segment is not computable).
However, it is interesting to mention that this example does not mean that the computability of the boundary
is necessary for the computability of the entire manifold. By \cite{mi:mib},
there exists a computable arc in $\mathbb{R}^{2} $ with noncomputable endpoints, hence
the computability of 1-manifold with boundary does not imply the computability
of its boundary.

Regarding the computability of a manifold, we can notice that this
does not mean that the manifold can be parameterized by a computable function.
Namely, by \cite{mi:mib}, there exists a computable arc $S$ in $\mathbb{R}^{2} $
with computable endpoints such that there exists no
computable
bijection $f:[0,1]\rightarrow S$.

In Section \ref{prelim} we give necessary  definitions and some basic facts.
In Section \ref{sect-3} we define semi-c.c.b.\ sets. In Section \ref{sect-4} we
introduce chains and we develop certain techniques which we will need later.
In Section \ref{sect-5} we prove that $S$ is computably enumerable if $S\cup F$ is semi-c.c.b., where
$F$ is a closed set disjoint with $S$ and $S$ is a topological ray with computable
endpoint (Theorem \ref{top-ray}). In Section \ref{sect-6} we prove the same
under assumption that $S$ is a topological line (Theorem \ref{top-line}).
Finally, in Section \ref{sect-7} we get that each semi-c.c.b.\ 1-manifold with boundary which has
finitely many components is computable if its boundary is semi-c.c.b.\ (Theorem \ref{glavni}).
 This in particular
means that each semi-c.c.b.\ (boundaryless) 1-manifold  which has
finitely many components is computable. In Section \ref{sect-7} we will actually prove
this: if $M$ is a 1-manifold with boundary and if both $M$ and $\partial M$ are
semi-c.c.b., then each component of $M$ is computably enumerable (Theorem \ref{komp-ce}).

Let us mention  that the uniform version of the result for 1-manifolds
(Theorem \ref{glavni})
 does not hold in general. Namely, by Example 7 in \cite{zi:mib}, there exists a
 sequence $(S_{i} )$ of topological circles in $\mathbb{R}^{2}$
 such that $S_{i} $ is uniformly semi-computable, but
 not uniformly computable. Moreover, each $S_{i} $ is contained in the compact set $[0,1]\times [0,1]$.

\section{Basic notions and techniques}\label{prelim}

If $X$ is a set, let $\mathcal{P}(X)$ denote the set of all
subsets of $X$.

For $m\in \mathbb{N}$ let  $\mathbb{N}_{m} =\{0,\dots ,m\}$. For
$n\geq 1$ let $$\mathbb{N}_{m} ^{n} =\{(x_{1} ,\dots ,x_{n} )\mid
x_{1} ,\dots ,x_{n}\in \mathbb{N}_{m} \}.$$

We say that a function $\Phi :\mathbb{N}^{k} \rightarrow
\mathcal{P}(\mathbb{N}^{n} )$ is \textbf{computable finitely valued} or \textbf{c.f.v.} if the
function $\overline{\Phi }:\mathbb{N}^{k+n}\rightarrow \mathbb{N}$
defined by
$$\overline{\Phi }(x,y)=\chi _{\Phi (x)}(y),$$ $x\in \mathbb{N}^{k} ,$ $y\in \mathbb{N}^{n}$
is computable (i.e.\ recursive), where $\chi _{S}:\mathbb{N}^{n}
\rightarrow \{0,1\}$ denotes the characteristic function of
$S\subseteq \mathbb{N}^{n} $, and if there exists a computable function
$\varphi :\mathbb{N}^{k} \rightarrow \mathbb{N}$ such that $$\Phi
(x)\subseteq \mathbb{N}_{\varphi (x)}^{n} $$  for all $x\in
\mathbb{N}^{k}$.

\begin{prop} \label{p1}\hfill
\begin{enumerate}
\item If $\Phi ,\Psi:\mathbb{N}^{k} \rightarrow
\mathcal{P}(\mathbb{N}^{n} )$ are c.f.v.\ functions, then the
function $\mathbb{N}^{k} \rightarrow \mathcal{P}(\mathbb{N}^{n}
)$, $x\mapsto \Phi (x)\cup \Psi (x)$ is c.f.v.

\item If $\Phi ,\Psi:\mathbb{N}^{k} \rightarrow
\mathcal{P}(\mathbb{N}^{n} )$ are c.f.v.\ functions, then the sets
$\{x\in \mathbb{N}^{k} \mid \Phi (x)=\Psi (x)\}$ and $\{x\in
\mathbb{N}^{k} \mid \Phi (x)\subseteq \Psi (x)\}$ are decidable.

\item Let $\Phi :\mathbb{N}^{k} \rightarrow
\mathcal{P}(\mathbb{N}^{n} )$ and $\Psi :\mathbb{N}^{n}\rightarrow
\mathcal{P}(\mathbb{N}^{m} )$ be c.f.v.\ functions. Let $\Lambda
:\mathbb{N}^{k} \rightarrow \mathcal{P}(\mathbb{N}^{m} )$ be
defined by $$\Lambda (x)=\bigcup_{z\in \Phi (x)}\Psi  (z),$$ $x\in
\mathbb{N}^{k} $. Then $\Lambda  $ is a c.f.v.\ function.

\item Let $\Phi :\mathbb{N}^{k} \rightarrow
\mathcal{P}(\mathbb{N}^{n} )$ be c.f.v$.$ and let $T\subseteq
\mathbb{N}^{n} $ be c.e. Then the set $S=\{x\in \mathbb{N}^{k}
\mid \Phi (x)\subseteq T\}$ is c.e. \qed
\end{enumerate}
\end{prop}

\subsection{Computable metric spaces}

A function $F:\mathbb{N}^{k} \rightarrow \mathbb{Q}$ is called
\textbf{computable} if there exist computable functions
$a,b,c:\mathbb{N}^{k}\rightarrow \mathbb{N}$ such that
$$F(x)=(-1)^{c(x)}\frac{a(x)}{b(x)+1}$$ for each $x\in
\mathbb{N}^{k} $. A number $x\in \mathbb{R}$ is said to be
\textbf{computable} if there exists a computable function
$g:\mathbb{N}\rightarrow \mathbb{Q}$ such that $|x-g(i)|<2^{-i}$
for each $i\in \mathbb{N}$ \cite{turing}.

By a \textbf{computable} function $\mathbb{N} ^{k} \rightarrow
\mathbb{R}$ we mean a function $f:\mathbb{N} ^{k} \rightarrow
\mathbb{R}$ for which there exists a computable function
$F:\mathbb{N}^{k+1}\rightarrow \mathbb{Q}$ such that
$$|f(x)-F(x,i)|<2^{-i}$$ for all $x\in \mathbb{N}^{k}$ and $i\in
\mathbb{N}$.

\begin{prop} \label{NuR}\hfill
\begin{enumerate}

\item If $f,g:\mathbb{N}^{k} \rightarrow \mathbb{R}$ are
computable, then $f+g,f-g:\mathbb{N}^{k} \rightarrow
\mathbb{R}$ are computable.

\item If $f,g:\mathbb{N}^{k} \rightarrow \mathbb{R}$ are
computable functions, then the set $\{x\in \mathbb{N}^{k} \mid
f(x)>g(x)\}$ is c.e.

\qed

\end{enumerate}
\end{prop}

A tuple $(X,d,\alpha )$ is said to be a \textbf{computable metric
space} if $(X,d)$ is a metric space and $\alpha :\mathbb{N}
\rightarrow X$ is a sequence dense in $(X,d)$ (i.e.\ a sequence
 the range of which is dense in $(X,d)$) such that the function
$\mathbb{N} ^{2}\rightarrow \mathbb{R} $, $$(i,j)\mapsto
d(\alpha_{i} ,\alpha_{j} )$$ is computable (we use notation
$\alpha =(\alpha _{i} )$).

If  $(X,d,\alpha )$ is a computable metric space, then a sequence
$(x_{i} )$ in $X$ is said to be \textbf{computable} in
$(X,d,\alpha )$ if there exists a computable function
$F:\mathbb{N} ^{2}\rightarrow \mathbb{N} $ such that $$d(x_{i}
,\alpha _{F(i,k)})<2^{-k}$$ for all $i,k\in \mathbb{N} $.
A point
$a \in X$ is said to be \textbf{computable} in $(X,d,\alpha )$ if
there exists a computable function $f:\mathbb{N}\rightarrow
\mathbb{N}$ such that $d(a,\alpha _{f(k)})<2^{-k}$ for each $k\in
\mathbb{N}$.

The points   $\alpha _{0}$, $\alpha _{1} $, $\dots$  are called
\textbf{rational points}. If $i\in \mathbb{N}$ and $q\in
\mathbb{Q}$, $q>0$, then we say that  $B(\alpha _{i} ,q)$ is an
(open) \textbf{rational ball}. Here, for $x\in X$ and $r>0$, we
denote by $B(x,r)$ the open ball of radius $r$ centered at $x$,
i.e$.$ $B(x,r)=\{y\in X\mid d(x,y)<r\}$. By $\widehat{B}(x,r)$
(for $x\in X$ and $r\geq 0$) we will denote the corresponding
closed ball $\{y\in X\mid d(x,y)\leq r\}$.

If $B_{1} ,\dots ,B_{n} $, $n\geq 1$, are open rational balls,
then the union $B_{1}\cup \dots \cup B_{n} $
 will be called a \textbf{rational open set}.

\begin{exa} \label{ex1} If $\alpha :\mathbb{N} \rightarrow
\mathbb{R}^{n}  $ is a computable function (in the sense that the
component functions of $\alpha $ are computable) whose image is
dense in $\mathbb{R}^{n} $ and $d$ is the Euclidean metric on
$\mathbb{R}$, then $(\mathbb{R}^{n}  , d, \alpha )$ is a
computable metric space. A sequence $(x_{i} )$ is computable in
this computable metric space if and only if $(x_{i} )$ is a
computable sequence in $\mathbb{R}^{n} $  and $(x_{1} ,\dots
,x_{n} )\in \mathbb{R}^{n} $ is a computable point in this space
if and only if $x_{1} $,\dots ,$x_{n} $ are computable numbers.
\end{exa}

\subsection{Effective enumerations}

Let $(X,d,\alpha )$ be a computable metric space. Let
$q:\mathbb{N}\rightarrow \mathbb{Q}$ be some fixed computable
function whose image is $\mathbb{Q}\cap \langle 0,\infty\rangle $
and let $\tau_{1}   ,\tau_{2}   :\mathbb{N}\rightarrow \mathbb{N}$
be some fixed computable functions such that $\{(\tau _{1} (i),
\tau _{2} (i))\mid i\in \mathbb{N}\}=\mathbb{N}^{2}.$ Let
$(\lambda _{i} )_{i\in \mathbb{N}}$ be the sequence of points in
$X$ defined by $\lambda _{i} =\alpha _{\tau _{1} (i)}$ and let
$(\rho _{i} )_{i\in \mathbb{N}}$ be the sequence of rational
numbers defined by $\rho _{i} =q_{\tau _{2} (i)}$. For $i\in
\mathbb{N}$ we define
$$I_{i}=B(\lambda_{i}   ,\rho_{i}  ),~\widehat{I}_{i}=\widehat{B}(\lambda_{i}   ,\rho_{i}  ).$$ The sequences $(I_{i} )$ and $(\widehat{I}_{i})$
represent  effective
enumerations of all open rational balls and all closed rational
balls.

Let $\sigma :\mathbb{N}^{2}\rightarrow \mathbb{N}$ and
$\eta:\mathbb{N}\rightarrow \mathbb{N}$ be some fixed computable
functions with the following property: $\{(\sigma (j,0),\dots
,\sigma (j,\eta(j)))\mid j\in \mathbb{N}\}$ is the set of all
finite sequences in $\mathbb{N}$ (excluding the empty sequence),
i.e. the set $\{(a_{0} ,\dots ,a_{n} )\mid n\in  \mathbb{N},~a_{0}
,\dots ,a_{n} \in \mathbb{N}\}$.  We use the following notation:
$(j)_{i}$ instead of $\sigma (j,i)$ and $\overline{j}$ instead of
$\eta(j).$  Hence $$\{((j)_{0} ,\dots ,(j)_{\overline{j}})\mid
j\in \mathbb{N}\}$$ is the set of all finite sequences in
$\mathbb{N}.$ For $j\in \mathbb{N}$ let
\begin{equation}\label{p2-eq}
[j]=\{(j)_{i} \mid 0\leq i\leq \overline{j}\}.
\end{equation}

For $j\in \mathbb{N}$ we define
$$J_{j}=\bigcup_{i\in [j]}I_{i}.$$
Then  $(J_{j}) $ is an effective enumeration of all rational open
sets.

Note that the function $\mathbb{N}\rightarrow \mathcal{P}(\mathbb{N})$, $j\mapsto [j]$ is c.f.v.\ (Proposition \ref{p1}(3)).
Also note that any finite nonempty subset of $\mathbb{N}$ equals $[j]$ for some $j\in \mathbb{N}$.
\begin{cor} \label{c.f.v.-repr}
Let $\Phi :\mathbb{N}^{k} \rightarrow \mathcal{P}(\mathbb{N})$ be a c.f.v.\ function such that
$\Phi (x)\neq\emptyset $ for each $x\in \mathbb{N}^{k}$. Then there exists a computable function $\varphi :\mathbb{N}^{k} \rightarrow \mathbb{N}$ such that
$\Phi (x)=[\varphi (x)]$ for each $x\in \mathbb{N}^{k} $.
\end{cor}
\proof For each $x\in \mathbb{N}^{k} $ there exists $j\in \mathbb{N}$ such that $\Phi (x)=[j]$. Since the set of all
$(x,l)$, $x\in \mathbb{N}^{k} $, $l\in \mathbb{N}$, for which $\Phi (x)=[j]$ holds is decidable by Proposition \ref{p1}(2), for each
$x\in \mathbb{N}^{k} $ we can effectively find $j\in \mathbb{N}$ such that $\Phi (x)=[j]$. \qed

\subsection{Formal properties}

In Euclidean space $\mathbb{R}^{n}$ we can effectively calculate
the diameter of the finite union of rational balls. However, in a
general computable metric space the function
$\mathbb{N}\rightarrow \mathbb{R}$, $j\mapsto \diam(J_{j} )$, need
not be computable. This is the reason that we are going to use the
notion of the formal diameter.

Let $(X,d)$ be a metric space and $x_{0},\dots ,x_{k} \in X,$
$r_{0} ,\dots ,r_{k} \in \mathbb{R}_{+}.$ The \textbf{formal
diameter} associated to the finite sequence $(x_{0},r_{0} ),\dots
,(x_{k},r_{k} ) $ is the number $D\in \mathbb{R}$ defined by
$$D=\max_{0\leq v,w\leq k }d(x_{v} ,x_{w} )+2\max_{0\leq v\leq
k}r_{v}.$$ It follows from this definition that $\diam(B(x_{0}
,r_{0} )\cup \dots \cup B(x_{k} ,r_{k} ))\leq D$.

Let $(X,d,\alpha )$ be a computable metric space. We define the
function $\fdiam:\mathbb{N}\rightarrow \mathbb{R}$ in the
following way. For $j\in \mathbb{N}$ the number $\fdiam(j)$ is the
formal diameter associated to the finite sequence
$$\left(\lambda _{(j)_{0}} , \rho _{(j)_{0}}\right),\dots ,\left(\lambda
_{(j)_{\overline{j}}} , \rho  _{(j)_{\overline{j}}}\right).$$
 Clearly $\diam(J_{j}
)\leq \fdiam(j)$ for each $j\in \mathbb{N}$.

Let $i,j\in \mathbb{N}$. We say that $I_{i} $ and $I_{j} $ are
\textbf{formally disjoint} if $$d(\lambda_{i} ,\lambda_{j}
)>\rho_{i} +\rho_{j} .$$ Note that we define this as a relation
between the numbers $i$ and $j$, not the sets $I_{i} $ and $I_{j}
$.

Let $i,j\in \mathbb{N}$. We say that $J_{i} $ and $J_{j} $ are
\textbf{formally disjoint} if $I_{k}$ and $I_{l} $ are formally
disjoint for all $k\in [i]$ and $l\in [j]$. Clearly, if $J_{i} $
and $J_{j} $ are formally disjoint, then $J_{i} \cap J_{j}
=\emptyset $.

We will also say that $I_{i}$ and $J_{j} $ are
\textbf{formally disjoint} if $I_{i}$ and $I_{l} $ are formally
disjoint for each $l\in [j]$. Note that formal disjointness of $I_{i}$ and $J_{j} $
implies $\widehat{I}_{i}\cap J_{j} =\emptyset $.

Let $i,m\in \mathbb{N}$ and $a\in X$. We say that $I_{i} $ is
\textbf{formally contained} in $B(a,m)$ and write $I_{i} \subseteq
_{F}B(a,m)$ if $d(\lambda _{i} ,a)+\rho _{i} <m$ (again, this as a
relation between $i$, $a$ and $m$, not between $I_{i} $ and
$B(a,m)$). Clearly, if $I_{i} \subseteq _{F}B(a,m)$, then $I_{i}
\subseteq B(a,m)$. For $j\in \mathbb{N}$ we write $$J_{j}
\subseteq _{F} B(a,m)$$ if $I_{i} \subseteq _{F}B(a,m)$ for each
$i\in [j]$. If $J_{j} \subseteq _{F} B(a,m)$, then $J_{j}
\subseteq B(a,m)$.

In the same way we define that $I_{i} $ is formally contained in $I_{m} $
($I_{i} \subseteq _{F}I_{m}$)
and that $J_{j} $ is formally contained in $I_{m}$ ($J_{j} \subseteq _{F}I_{m} $).

\begin{prop} \label{fdiam-FD}
\begin{enumerate}
\item The function $\fdiam:\mathbb{N}\rightarrow \mathbb{R}$ is
computable.

\item The sets $\{(i,j)\in \mathbb{N}^{2}\mid I_{i} $ and $J_{j} $
are formally disjoint$\}$ and $\{(i,j)\in \mathbb{N}^{2}\mid J_{i} $ and $J_{j} $
are formally disjoint$\}$ are c.e.

\item If $a$ is a computable point, then the set $\{(j,m)\mid
J_{j} \subseteq _{F}B(a,m)\}$ is c.e.

\item The set $\{(j,m)\in \mathbb{N}^{2}\mid J_{j} \subseteq _{F} I_{m} \}$
is c.e.
 \qed
\end{enumerate}
\end{prop}
\proof For (1) and (2) see \cite[Proposition 2.4]{lmcs:mnf}.
 Let us prove (3). Let $$\Omega =\{(j,m)\mid J_{j}
\subseteq _{F}B(a,m)\}\mbox{ and }\Gamma  =\{(i,m)\mid I_{i}
\subseteq _{F}B(a,m)\}.$$ Let $\Phi :\mathbb{N}^{2}\rightarrow
\mathcal{P}(\mathbb{N}^{2})$ be defined by $\Phi (j,m)=[j]\times
\{m\}$. Then
$$(j,m)\in \Omega \Leftrightarrow \Phi (j,m)\subseteq \Gamma .$$
We have that $\Phi $ is c.f.v. So if we prove that $\Gamma $ is c.e., we will
have that $\Omega $ is c.e.\ (Proposition \ref{p1}). However, the fact that $\Gamma $ is
c.e.\ follows from Proposition \ref{NuR} since
$$(i,m)\in \Gamma \Leftrightarrow d(\lambda _{i} ,a)+\rho _{i} <m.$$
In the same way we get (4). \qed

The following simple lemma will be very useful to us later.

\begin{lem} \label{fdiam-fdisj}
Let $m\in \mathbb{N}$ and let $x\in I_{m} $. Then there exists
$\varepsilon >0$ with the following property: if $j\in \mathbb{N}$ is such that
$x\in J_{j} $ and $\fdiam(j)<\varepsilon $, then $J_{j} \subseteq _{F}I_{m} $.
\end{lem}
\proof We have $d(\lambda _{m} ,x)<\rho _{m} $ and therefore
there exists $r >0$ such that $$d(\lambda _{m} ,x)+r<\rho _{m} .$$

Let $\varepsilon= \frac{r}{2}$. Suppose $j\in \mathbb{N}$ is such that
$x\in J_{j} $ and $\fdiam(j)<\varepsilon $.
Let $i\in [j]$. Then $\rho _{i} <\fdiam(j)<\varepsilon $
and $d(x,\lambda _{i} )\leq \diam(J_{j} )\leq \fdiam(j)<\varepsilon $. We have
$$d(\lambda _{m} ,\lambda _{i} )+\rho _{i} \leq d(\lambda _{m} ,x) + d(x,\lambda _{i} )+\rho _{i}< d(\lambda _{m} ,x) + 2\varepsilon =d(\lambda _{m} ,x) + r<\rho _{m} .$$
So $d(\lambda _{m} ,\lambda _{i} )+\rho _{i}<\rho _{m} $ and $I_{i} \subseteq _{F}I_{m} $.
Hence $J_{j} \subseteq _{F}I_{m}$ .\qed

\subsection{Computable sets} Let $(X,d,\alpha )$ be a computable metric space.
A closed subset $S$ of $(X,d)$ is said to be \textbf{computably
enumerable} in $(X,d,\alpha )$ if
$$\{i\in \mathbb{N}\mid S \cap I_{i} \neq\emptyset \}$$ is a
c.e$.$ subset of $\mathbb{N}.$ A closed subset $S$ of $(X,d)$ is
said to be \textbf{co-computably enumerable} in $(X,d,\alpha )$ if
there exists a computable function $f:\mathbb{N}\rightarrow
\mathbb{N}$ such that
$$X\setminus S=\bigcup _{i\in \mathbb{N}}I_{f(i)}.$$  We say that $S$ is a \textbf{computable
set} in $(X,d,\alpha )$ if $S$ is  a computably enumerable and a
co-computably enumerable set (\cite{bp:mib,we}).

Let $(X,d,\alpha )$ be a computable metric space. We say that $K$
is a \textbf{semi-computable compact set} in $(X,d,\alpha )$ if
$K$ is a compact set in $(X,d)$ and if the set $\{j\in
\mathbb{N}\mid K\subseteq J_{j} \}$  is c.e. We say that $K$ is a
\textbf{computable compact set} if  $K$ is a semi-computable
compact set and $K$ is computably enumerable.

\section{Ambient space and c.c.b.\  sets} \label{sect-3}

A computable metric space $(X,d,\alpha )$ has the
\textbf{effective covering property} if the set $$\{(i,j)\in
\mathbb{N}^{2}\mid \widehat{I}_{i}\subseteq J_{j}\}$$ is
computably enumerable (\cite{bp:mib}). Euclidean space
$\mathbb{R}^{n} $ (example \ref{ex1}) has the effective covering
property (see e.g.\ \cite{zi:mib}).

A computable metric space which has the effective covering
property and in which each closed ball is compact has a property
which turns out to be important if we want to get that some set is
computable. The property is this: if $S$ is compact and co-c.e.,
then we can effectively enumerate all rational open sets which
cover $S$. In other words, if a compact set is co-c.e., then it is
semi-computable compact.

We have mentioned the result from \cite{lmcs:mnf} regarding the
computability of co-c.e.\ compact manifolds. In \cite{lmcs:mnf}
the following is proved:
\begin{fact} \label{fact-1}
If a computable metric space  has the effective covering property
and compact closed balls, then each co-c.e.\ compact manifold in
this space with computable boundary is computable.
\end{fact}
However, this result is just a consequence of the following result
which is also proved in \cite{lmcs:mnf}:
\begin{fact} \label{fact-2}
In any computable metric space any compact manifold which is
semi-computable compact and whose boundary is computable compact
is computable compact.
\end{fact}
Note that in Fact \ref{fact-2} we have the stronger assumptions
(and the stronger conclusion) on the sets, but there are no
assumptions on the ambient space. Since the notions of a co-c.e.\
set and a semi-computable compact set coincide for compact sets in
computable metric space with the effective covering property and
compact closed balls, the Fact \ref{fact-2} is clearly a
generalization of Fact \ref{fact-1}.

In this paper we examine 1-manifolds, the sets which are not
compact in general. We will have the result that if a 1-manifold
with finitely many components is co-c.e.\ and its boundary is
computable, then this manifold is computable. However, we will
need for this result the assumption that the ambient space has the
effective covering property and compact closed balls.   We would
like to find some analogue of the notion of a semi-computable set
for noncompact sets so that, in the same manner as in the case of
compact manifolds, we can remove the assumptions on the computable metric
space. Of course, we want that the new result which holds in
general computable metric spaces be the generalization of the
previous result for co-c.e.\ sets in the computable metric spaces
with effective covering property and compact closed balls. And
this will be true if this analogue of semi-computability coincides
with the the notion of a co-c.e.\ set in these special computable
metric spaces.

That a set $S$ is semi-computable compact means that we can
effectively enumerate all rational open sets which cover $S$. The
idea for a generalization of this notion is to observe a set $S$
which may not be compact, but such that the intersection $S\cap B$
is compact for each closed ball $B$ in the ambient space and
furthermore such that we can effectively (and uniformly) enumerate
all rational open sets which cover $S\cap B$ for each closed ball
$B$.

Let $(X,d,\alpha )$ be a computable metric space. Let $S\subseteq
X$. We say that $S$ is \textbf{c.c.b}.\ (or \textbf{computable compact on closed balls}) if the following holds:
\begin{enumerate}
\item[(1)] $S\cap \ \widehat{B}(x,r)$ is a compact set for all
$x\in X$ and $r>0$;

\item[(2)] the set $\{(i,j)\in  \mathbb{N}^{2}\mid \widehat{I}_{i}
\cap S\subseteq J_{j} \}$ is c.e.;

\item[(3)] $S$ is computably enumerable.
\end{enumerate}

If $S$ is a set which satisfies conditions (1) and (2), then we
will say that $S$ is \textbf{semi-c.c.b}. Hence $S$ is c.c.b.\  if
and only if $S$ is semi-c.c.b.\  and computably enumerable. Note
that semi-c.c.b.\ sets (and c.c.b.\ sets) are closed (this follows
from (1)).

Let $(X,d,\alpha )$ be a computable metric space. Then $X$ is semi-c.c.b.\  in $(X,d,\alpha )$ if and
only if $(X,d,\alpha )$ has compact closed balls and the effective
covering property. For example, $\mathbb{R}^{n} $ is semi-c.c.b.\  (and also c.c.b.) in the computable
metric space from Example \ref{ex1}. Hence semi-c.c.b.\  sets (and also c.c.b.\  sets) need not be compact.

On the other hand, we now show that each semi-computable compact set is semi-c.c.b. In other words, the notion of a semi-c.c.b.\  set generalizes the
notion of a semi-computable compact set.
\begin{prop} \label{scc-sccb}
Let $(X,d,\alpha )$ be a computable metric space. Let $S$ be a
semi-computable compact set in this space. Then $S$ is semi-c.c.b.
\end{prop}
\proof We have to show that the set $\{(i,j)\in \mathbb{N}^{2}\mid
\widehat{I}_{i} \cap S\subseteq J_{j} \}$ is c.e.

Suppose $i,j\in \mathbb{N}$ are such that $\widehat{I}_{i} \cap
S\subseteq J_{j}$. Let $x\in S\setminus J_{j} $. Then $x\notin
\widehat{I}_{i}$ and therefore there exists some $k_{x}\in
\mathbb{N}$ such that $x\in I_{k_{x}}$ and such that $I_{i} $ and
$I_{k_{x}}$ are formally disjoint. The set $S\setminus J_{j} $ is
closed, hence compact (since $S$ is compact) and this implies that
there exist $n\in \mathbb{N}$ and $x_{0} ,\dots ,x_{n} \in
S\setminus J_{j} $ such that $S\setminus J_{j} \subseteq
I_{k_{x_{0} }}\cup \dots \cup I_{k_{x_{n} }}$. It follows
$$S \subseteq J_{j} \cup I_{k_{x_{0} }}\cup \dots \cup
I_{k_{x_{n} }}.$$
Therefore, there exists $l\in \mathbb{N}$ such that
\begin{equation}\label{ccb-1}
S\subseteq J_{j} \cup J_{l} \mbox{ and }I_{i} \mbox{ and }J_{l} \mbox{ are formally disjoint.}
\end{equation}

On the other hand, suppose that (\ref{ccb-1}) holds for some $i,j,l\in \mathbb{N}$. Then $\widehat{I}_{i} \cap J_{l} =\emptyset $ and therefore
$\widehat{I}_{i} \cap S\subseteq J_{j} $. Hence we have the following conclusion:
$\widehat{I}_{i} \cap S\subseteq J_{j} $ if and only if there exists $l\in \mathbb{N}$ such that
(\ref{ccb-1}) holds.

The function $\mathbb{N}^{2}\rightarrow \mathcal{P}(\mathbb{N})$, $(j,l)\mapsto [j]\cup [l]$ is c.f.v., therefore by Corollary \ref{c.f.v.-repr}
there exists a computable function $\varphi :\mathbb{N}\rightarrow \mathbb{N}$ such that $[j]\cup [l]=[\varphi (j,l)]$ for all $j,l\in \mathbb{N}$.
Hence $J_{j} \cup J_{l} =J_{\varphi (j,l)}$ for all $j,l\in \mathbb{N}$ and using the fact that $S$ is semi-computable we conclude
that the set $\{(j,l)\in \mathbb{N}^{2}\mid S\subseteq J_{j} \cup J_{l} \}$ is c.e. This implies that the set of
all $(i,j,l)\in \mathbb{N}^{3}$ such  that (\ref{ccb-1}) holds is c.e.\ (Proposition \ref{fdiam-FD}) and we conclude that the set
$\{(i,j)\in \mathbb{N}^{2}\mid \widehat{I}_{i} \cap S\subseteq J_{j}\}$ is c.e. \qed

Note that semi-computable compact sets are exactly those semi-c.c.b.\  sets which are compact.
(If $S$ is a compact set, then  $S\subseteq \widehat{I}_{i_{0} }$ for some $i_{0} \in \mathbb{N}$, so if $S$ is semi-c.c.b,
the set $\{j\in \mathbb{N}\mid \widehat{I}_{i_{0} }\cap S\subseteq J_{j} \}$
is c.e. This set clearly equals $\{j\in \mathbb{N}\mid S\subseteq J_{j} \}$.)

Now we show that semi-c.c.b.\  sets are co-c.e. First we have the following property of semi-c.c.b.\  sets.
\begin{prop} \label{ccb-disj}
Let $(X,d,\alpha )$ be a computable metric space. Let $S$ be a
semi-c.c.b.\  set. Then the set
$$\Omega =\{i\in \mathbb{N}\mid \widehat{I}_{i} \cap S=\emptyset \}$$
is c.e.
\end{prop}
\proof We may assume that $S\neq \emptyset $.
Let $\Gamma =\{(i,j)\in \mathbb{N}^{2}\mid \widehat{I}_{i} \cap S\subseteq J_{j} \}$.

Suppose $i\in \Omega $. Then $\widehat{I}_{i} \cap S=\emptyset$ which implies $\widehat{I}_{i}\neq X$
and therefore there exists $j\in \mathbb{N}$ such that $I_{i} $ and $J_{j} $ are formally disjoint.
Clearly $\widehat{I}_{i} \cap S\subseteq J_{j}$, hence $(i,j)\in \Gamma $.

Conversely, let us take  $j\in \mathbb{N}$ such that $I_{i} $ and
$J_{j} $ are formally disjoint and $(i,j)\in \Gamma $. Then
$\widehat{I}_{i} \cap J_{j}=\emptyset $. But we have
$\widehat{I}_{i} \cap S\subseteq J_{j}$ and this can only be true
if $\widehat{I}_{i} \cap S=\emptyset $. Hence $i\in \Omega $.

We have the following conclusion:
$$i\in \Omega \Longleftrightarrow \mbox{ there exists } j\in \mathbb{N} \mbox{ such that }(i,j)\in \Gamma\mbox{ and }I_{i}\mbox{ and }J_{j}\mbox{ are formally disjoint}.$$
The fact that $\Gamma $ is c.e.\ and Proposition \ref{fdiam-FD} imply that $\Omega $ is c.e. \qed

\begin{prop}
Let $(X,d,\alpha )$ be a computable metric space. Let $S$ be a
semi-c.c.b.\  set. Then $S$ is co-c.e.
\end{prop}
\proof Let $x\in X\setminus S$. Since $S$ is closed, we have $B(x,r)\subseteq X\setminus S$ for
some $r>0$. Take a rational point $a$ and a positive rational number $\lambda $ so that $\lambda <\frac{r}{2}$
and $x\in B(a,\lambda )$. Then $\widehat{B}(a,\lambda )\cap S=\emptyset $. The conclusion is this: for each point $x\in X\setminus S$
there exists $i\in \mathbb{N}$ such that $x\in I_{i} $ and $\widehat{I}_{i} \cap S=\emptyset $.

Let $\Omega =\{i\in \mathbb{N}\mid \widehat{I}_{i} \cap S=\emptyset \}$. It follows from the previous
fact that $$X\setminus S=\bigcup_{i\in \Omega }I_{i} .$$
However  $\Omega $ is c.e.\ by Proposition \ref{ccb-disj}
and this means that $S$ is co-c.e.  \qed

In general, a co-c.e.\ set need not be semi-c.c.b, even if it is compact. Moreover, even a singleton set need not be
semi-computable compact if it is co-c.e. To see this, note first the following: if $(X,d,\alpha )$ is a computable metric
space and $x\in X$ such that $\{x\}$ is semi-computable compact, then $x$ is
a computable point. Namely, for each $k\in \mathbb{N}$ there exists $j\in \mathbb{N}$ such that
\begin{equation}\label{ccb-2}
\{x\}\subseteq J_{j}\mbox{ and }\fdiam(j)<2^{-k}.
\end{equation}
Since the set of all $(k,l)\in \mathbb{N}^{2}$ for which (\ref{ccb-2}) holds
is c.e., there exists a computable function $\varphi :\mathbb{N}\rightarrow \mathbb{N}$ such that
(\ref{ccb-2}) holds for each $k\in \mathbb{N}$ and $l=\varphi (k)$. Recall $J_{j} =I_{(j)_{0} }\cup \dots \cup I_{(j)_{\overline{j}}}$
and $I_{(j)_{0} }=B(\lambda_{(j)_{0}} ,\rho_{(j)_{0}} )$. So we have $d(x,\lambda _{(j)_{0} })<2^{-k}$ for each $k\in \mathbb{N}$ and $x$ is computable point.

By Example 3.2.\ in \cite{glasnik} there exists a computable metric space $(X,d,\alpha )$ and a point
$x\in X$ such that $\{x\}$ is co-c.e., but $x$ is not a computable point. Therefore $\{x\}$ is not a semi-computable compact set.

We have mentioned that in a computable metric space which has the effective covering
property and compact closed balls a set is semi-computable compact if and only if it is compact and co-c.e.
Now we prove a more general result.

\begin{prop} \label{prop-co-c.e.-semi}
Let $(X,d,\alpha )$ be a computable metric space. Suppose
$(X,d,\alpha )$ has the effective covering property and compact
closed balls. Let $S\subseteq X$. Then $S$ is co-c.e.\ if and only
if $S$ is semi-c.c.b.
\end{prop}
\proof We have to prove that if $S$ is co-c.e., then $S$ is semi-c.c.b.
Suppose $S$ is co-c.e. It is easy to conclude that then there exists a computable function $f:\mathbb{N}\rightarrow \mathbb{N}$ such that
$J_{f(k)}\subseteq
J_{f(k+1)}$ for each $k\in \mathbb{N}$ and $$X\setminus
S=\bigcup_{k\in \mathbb{N}}J_{f(k)}.$$
Let $i,j\in \mathbb{N}$ and suppose that $\widehat{I}_{i} \cap S\subseteq J_{j} $.
It follows that the set $\widehat{I}_{i}\setminus J_{j} $ is contained in $X\setminus S$.
The set $\widehat{I}_{i}\setminus J_{j} $ is compact and therefore there exists $k\in \mathbb{N}$ such that
$\widehat{I}_{i}\setminus J_{j}\subseteq J_{f(k)} $ and consequently
\begin{equation}\label{ccb-3}
\widehat{I}_{i}\subseteq  J_{j}\cup  J_{f(k)} .
\end{equation}
On the other hand, if (\ref{ccb-3}) holds for some $i,j,k\in \mathbb{N}$,
then $\widehat{I}_{i}\cap S\subseteq  J_{j}$ (since $S\cap J_{f(k)}=\emptyset $).
Hence $\widehat{I}_{i} \cap S\subseteq J_{j} $ if and only if there exists $k\in \mathbb{N}$ such that
(\ref{ccb-3}) holds. The set of all $(i,j,k)\in \mathbb{N}^{3}$ such that
(\ref{ccb-3}) holds is c.e.\ (we can find a computable function $\varphi :\mathbb{N}^{2}\rightarrow \mathbb{N}$
such that $J_{j}\cup  J_{f(k)}=J_{\varphi (i,k)}$ for all $i,k\in \mathbb{N}$ as in the
proof of Proposition \ref{scc-sccb} and $(X,d,\alpha )$ has the effective
covering property). Therefore the set of all $(i,j)\in \mathbb{N}^{2}$
such that $\widehat{I}_{i} \cap S\subseteq J_{j} $ is c.e., which means that $S$ is semi-c.c.b. \qed

An immediate consequence of the previous proposition is the fact
that in a  computable metric space which has the effective covering
property and compact closed balls a set $S$ is computable (closed) if and only if $S$ is c.c.b.

\section{Chains} \label{lanci} \label{sect-4}

If $S$ is a semi-c.c.b.\  set in a computable metric space  and if
$a$ is a rational point, then for a given $n\in \mathbb{N}$ we can
effectively enumerate all rational open sets which contain $S \cap
\widehat{B}(a,n)$. In general, the problem is that we do not know,
for a given rational open set $U=B_{1}\cup \dots \cup B_{m}$ which
contains $S \cap \widehat{B}(a,n)$, which of these rational balls
$B_{1},\dots , B_{m}$ intersects $S$.

If we can somehow, for given $n\in \mathbb{N}$ and $\varepsilon
>0$, effectively find a rational open set $U=B_{1}\cup
\dots \cup B_{m}$ which contains $S\cap \widehat{B}(a,n)$, such
that each of the rational balls $B_{1},\dots , B_{m}$ has the
diameter less then $\varepsilon $ and such that each of these
balls  intersects $S$, then we have that $S$ is computable.
Namely, we only have to prove that $S$ is computably enumerable
(since $S$ is semi-c.c.b.\  by assumption). And if $i\in
\mathbb{N}$, then it is not hard to see that $I_{i} $ intersects
$S$ if and only if $I_{i} $ (formally) contains some of the balls
$B_{1},\dots ,B_{m}$ for some $n\in \mathbb{N}$ and $\varepsilon
>0$.

In order to effectively get, for given $n\in \mathbb{N}$ and
$\varepsilon >0$, such a rational open set $U$, we will use the
notion of a chain.

Let $X$ be a metric space. A finite sequence $C_{0} ,\dots ,C_{m}
$ of nonempty open subsets of $X$ is said to be a \textbf{chain}
in $X$ if $C_{i} \cap C_{j} =\emptyset$ for all $i,j\in \{0,\dots
,m\}$ such that $|i-j|> 1$ (see \cite{cv, na}). We say that $C_{i}
$ ($0\leq i\leq m$) is a \textbf{link} of the chain $C_{0} ,\dots
,C_{m} $. If $\mathcal{A}=(A_{0} ,\dots ,A_{m} )$ is a finite
sequence of nonempty bounded subsets of $X$, we define
$$\mesh(\mathcal{A})=\max_{0\leq i\leq m}\diam(A_{i} ).$$ If
$\varepsilon$ is a positive real number and $\mathcal{C}$ is a
chain, we say that $\mathcal{C}$ is an $\varepsilon -$chain if
$\mesh(\mathcal{C})<\varepsilon $.

Let $(X,d,\alpha )$  be a computable metric space. For $l\in
\mathbb{N}$ let $\mathcal{H}_{l} $ be the finite sequence of sets
$J_{(l)_{0}}, \dots, J_{(l)_{\overline{l}}}$. Furthermore, for
$j,p,q\in \mathbb{N}$ let $\mathcal{H}_{l}^{p\leq{}  q}  $ be the
finite sequence of sets $J_{(l)_{p}}, \dots, J_{(l)_{q}}$ if
$p\leq q$, otherwise let $\mathcal{H}_{l}^{p\leq{} q}$ denote the
empty sequence.  Clearly $\mathcal{H}_{l} =\mathcal{H}_{l}
^{0\leq{} \overline{l}}$.

Let the function $\fmesh:\mathbb{N}\rightarrow \mathbb{R}$ be
defined by
$$\fmesh(l)=\max_{0\leq j\leq \overline{l}}\fdiam((l)_{j}),$$
$l\in \mathbb{N}$.

Let $l\in \mathbb{N}$. We say that $\mathcal{H}_{l} $ is a
\textbf{formal chain} if $J_{(l)_{i} }$ and $J_{(l)_{j} }$ are
formally disjoint for all $i,j\in \{0,\dots ,\overline{l}\}$ such
that $|i-j|> 1$.

Let $a\in X$ and $l,p,q,m\in \mathbb{N}$. We say that
$\mathcal{H}_{l} ^{p\leq{} q}$ is \textbf{formally contained} in
$B(a,m)$ if $J_{(l)_{i} }\subseteq _{F}B(a,m)$ for each $i\in \mathbb{N}$ such that
$p\leq i\leq q$.

\begin{prop} \label{formal-chain}
Let $(X,d,\alpha )$ be a computable metric space.
\begin{enumerate}
\item The function $\fmesh:\mathbb{N}\rightarrow \mathbb{R}$ is
computable.

\item The set $\{l\in \mathbb{N}\mid \mathcal{H}_{l} $ is a formal
chain$\}$ is c.e.

\item If $a$ is a computable point, then the set $$\Gamma =\{(l,p,q,m)\mid
\mathcal{H}_{l} ^{p\leq{} q}\mbox{ formally contained in }B(a,m)\}$$
is c.e. \qed
\end{enumerate}
\end{prop}
\proof For (i) and (ii) see Proposition 5.4.\ in \cite{lmcs:mnf}.

For the proof of (iii),
let $\Omega =\{(j,m)\mid
J_{j} \subseteq _{F}B(a,m)\}$. By Proposition \ref{fdiam-FD} $\Omega $ is c.e. Let
$\Phi:\mathbb{N}^{4}\rightarrow \mathcal{P}(\mathbb{N}^{2})$ be defined by
$$\Phi (l,p,q,m)=\{((l)_{i},m)\mid p\leq i\leq q\}.$$ Then $\Phi $ is c.f.v.\ (Proposition \ref{p1}(3))
and $(l,p,q,m)\in \Gamma $ if and only if $\Phi (l,p,q,m)\subseteq \Omega $.
Now $\Gamma $ is c.e.\ by Proposition \ref{p1}(4). \qed

Let $j,l\in \mathbb{N}$, We say that $J_{j} $ and $\mathcal{H}_{l} $ are \textbf{formally disjoint}
if $J_{j} $ and $J_{i}$ are formally disjoint for each $i\in [l]$.
The following Lemma is an easy consequence of Proposition \ref{fdiam-FD}(ii).
\begin{lem} \label{Jj-disj-Hl}
Let $(X,d,\alpha )$ be a computable metric space. Then the set of
all $(j,l)\in \mathbb{N}^{2}$ such that $J_{j} $ and
$\mathcal{H}_{l} $ are formally disjoint is c.e. \qed
\end{lem}
In the similar way we define that $I_{i} $ is formally disjoint
with $\mathcal{H}_{l}^{p\leq q}$ and the  statement similar to
Lemma \ref{Jj-disj-Hl} also holds.

If $\mathcal{A}=(A_{0} ,\dots ,A_{m} )$ is a finite sequence of
sets, then by $\bigcup \mathcal{A}$ we denote the union $A_{0}
\cup \dots \cup A_{m} $. If $\mathcal{A}$ is the empty sequence, we take $\bigcup \mathcal{A}=\emptyset $.
Let $S$ be a set. We say that
$\mathcal{A}$ \textbf{covers} $S$ if $S\subseteq \bigcup
\mathcal{A}$.

\begin{lem} \label{lem-zeta}
Let $(X,d,\alpha )$ be a computable metric space. Then there exists
a computable function $\zeta :\mathbb{N}^{3}\rightarrow
\mathbb{N}$ such that
$$\bigcup \mathcal{H}_{l} ^{p\leq{} q}=J_{\zeta (l,p,q)}$$
for all $l,p,q\in \mathbb{N}$ such that $p\leq q$.
\end{lem}
\proof Let $l,p,q\in \mathbb{N}$ be such that $p\leq q$. We have
$$\bigcup \mathcal{H}_{l} ^{p\leq{} q}=\bigcup_{j=p}^{q} J_{(l)_{j} }=\bigcup_{j=p}^{q}\left(I_{((l)_{j})_{0}} \cup \dots \cup I_{((l)_{j})_{\overline{(l)_{j}}}}\right) .$$
Let $\Lambda :\mathbb{N}^{3}\rightarrow \mathcal{P}(\mathbb{N})$ be defined by
$$\Lambda (l,p,q)=\bigcup_{j=p\mbox{ or }p\leq j\leq q}\left\{((l)_{j})_{0}, \dots, ((l)_{j})_{\overline{(l)_{j}}}\right\}.$$
Then $\Lambda $ is c.f.v.\ by Proposition \ref{p1}(3). Clearly, we have
\begin{equation}\label{lem-zeta-1}
\bigcup \mathcal{H}_{l} ^{p\leq{} q}=\bigcup_{i\in \Lambda (l,p,q)}I_{i}.
\end{equation}
for all $l,p,q\in \mathbb{N}$ such that $p\leq q$.
Since $\Lambda (l,p,q)\neq \emptyset $ for all $l,p,q\in \mathbb{N}$ (condition $j=p$ in the definition of $\Lambda $
ensures this), there exists a computable function $\zeta :\mathbb{N}\rightarrow \mathbb{N}$ such that
$\Lambda (l,p,q)=[\zeta (l,p,q)]$ for all $l,p,q\in \mathbb{N}$. This means that
\begin{equation}\label{lem-zeta-2}
\bigcup_{i\in \Lambda (l,p,q)}I_{i} =J_{\zeta (l,p,q)}
\end{equation}
for all $l,p,q\in \mathbb{N}$. Comparing (\ref{lem-zeta-1}) and (\ref{lem-zeta-2})
we see that $\zeta $ is the desired function. \qed

\begin{prop} \label{co-c.e.-semicomp-H-l}
Let $(X,d,\alpha )$ be a computable metric space. Suppose
$S$ is a semi-c.c.b.\  set.
\begin{enumerate}
\item The set $$\Gamma =\left\{(i,l,p,q)\in \mathbb{N}^{4}\mid
\mathcal{H}_{l} ^{p\leq{} q}\mbox{ covers }S\cap \widehat{I}_{i}   \right\}$$
is c.e.
\item Let $a$ be a rational point. The sets
$$\Omega =\left\{(n,l,p,q)\in \mathbb{N}^{4}\mid
\mathcal{H}_{l} ^{p\leq{} q}\mbox{ covers }S\cap \widehat{B}(a,n)  \right\}$$
$$\Omega' =\left\{(n,l,p,q,u)\in \mathbb{N}^{5}\mid
S\cap \widehat{B}(a,n)\subseteq \bigcup\mathcal{H}_{l} ^{p\leq{} q}\cup J_{u}  \right\}$$
are
c.e.
\end{enumerate}
\end{prop}
\proof Let $\zeta $ be the function from Lemma \ref{lem-zeta}.

(i) For all $i,l,p,q\in \mathbb{N}$ we have
$$(i,l,p,q)\in \Gamma \Leftrightarrow (S\cap \widehat{I}_{i}\subseteq J_{\zeta (l,p,q)}\mbox{ and }p\leq q)\mbox{ or }(p>q\mbox{ and }S\cap \widehat{I}_{i}=\emptyset ).$$
That $\Gamma $ is c.e.\ as the union of two c.e.\ sets follows now from Proposition \ref{ccb-disj} and the fact that $S$ is semi-c.c.b.

(ii) Let us first notice that the set $\{j\in \mathbb{N}\mid a\in J_{j} \}$ is c.e. This follows from the fact that
$$a\in J_{j} \Leftrightarrow \exists i\in \mathbb{N}\mbox{ such that }a\in I_{i} \mbox{ and }i\in [j]$$
and $\{i\in \mathbb{N}\mid a\in I_{i} \}$ is c.e.\ since
$$a\in I_{i} \Leftrightarrow d(a,\lambda _{i} )<\rho _{i} $$
(we use here Proposition \ref{NuR}).
It follows easily now that the set $\{(l,p,q)\in \mathbb{N}^{3}\mid a\in \bigcup \mathcal{H}_{l} ^{p\leq{} q}\}$
is c.e.

Since $a$ is a rational point, there exists a computable function
$f:\mathbb{N}\rightarrow \mathbb{N}$ such that $\widehat{B}(a,n)=\widehat{I}_{f(n)}$ for
each $n\in \mathbb{N}$ such that $n\geq 1$. Note that $\widehat{B}(a,0)=\{a\}$.

Let us observe the case $a\in S$. Then we have
$$(n,l,p,q)\in \Omega \Leftrightarrow (\mathcal{H}_{l} ^{p\leq{} q}\mbox{ covers }S\cap \widehat{I}_{f(n)}\mbox{ and }n\geq 1)\mbox{ or }(a\in \bigcup \mathcal{H}_{l} ^{p\leq{} q}\mbox{ and }n=0).$$
The fact that $\Gamma $ is c.e.\ implies that $\Omega $ is c.e.

Let us now observe the case $a\notin S$. Then we have
$$(n,l,p,q)\in \Omega \Leftrightarrow (\mathcal{H}_{l} ^{p\leq{} q}\mbox{ covers }S\cap \widehat{I}_{f(n)}\mbox{ and }n\geq 1)\mbox{ or }n=0$$
and it follows that $\Omega $ is c.e. In the same way we get that $\Omega '$ is c.e. \qed

Suppose $(X,d)$ is a metric space and $S$ is an arc in this space
(a continuous injective image of the segment $[0,1]$). Then for
each $\varepsilon >0$ there exists an $\varepsilon -$ chain in
$(X,d)$ which covers $S$. We will need an effective version of
this fact.

Let $(X,d,\alpha )$ be a computable metric space. Let $A\subseteq X$, $j\in \mathbb{N}$ and $r\in \mathbb{R}$, $r>0$.
 We write $\langle A,j,\lambda\rangle$ to denote the following fact:

\[
A\subseteq J_{j}\mbox{ and }(I_{i}\cap A\not=\emptyset\mbox{ and }\rho_{i}<\lambda\mbox{ for each }i\in[j]).
\]
Note that $\langle A,j,\lambda\rangle$ and $\lambda \leq \lambda '$ implies $\langle A,j,\lambda'\rangle$.

\begin{prop} \label{luk-formlanac}
Let $(X,d,\alpha )$ be a computable metric space and let
$f:[0,r]\rightarrow X$ be a continuous injection, where $r>0$. Let
$\varepsilon
>0$. Then there exists $n_{0} \in \mathbb{N}$, $n_{0} \geq 1$, such that for each $n\geq n_{0} $
there exist numbers $j_{0} ,\dots ,j_{n-1} \in \mathbb{N}$ such
that
\begin{enumerate}
\item $\langle f([\frac{i}{n}r,\frac{i+1}{n}r]) ,j_{i},\varepsilon \rangle $ for each $i\in \{0,\dots ,n-1\}$;

\item $J_{j_{i} }$ and $J_{j_{i'} }$ are formally disjoint for all
$i,i'\in \{0,\dots ,n-1\}$ such that $|i-i'|>1$;

\item $\fdiam(j_{i})<\varepsilon $ for each $i\in \{0,\dots
,n-1\}$.
\end{enumerate}
\end{prop}

Before we prove this proposition, we need some facts.

\begin{lem}
\label{lem: lamba_td_Ji_Jj_formalno_disjunktni} Let $(X,d,\alpha)$
be a computable metric space. Let $A$ and $B$ be compact, nonempty
and disjoint subsets of $X$. Then
\begin{enumerate}
\item For each $\varepsilon>0$ there exists $j\in\mathbb{N}$ such that
$\langle A,j,\varepsilon\rangle$.
\item For each $\varepsilon>0$ there exists $\lambda>0$
such that $\lambda < \varepsilon$ and if $j,j'\in\mathbb{N}$ and $A'\subseteq A$ and $B'\subseteq B$ are such that
$$\langle A',j,\lambda \rangle\mbox{ and }\langle B',j',\lambda \rangle,$$
then $J_{j}$ and $J_{j'}$ are formally disjoint.
\end{enumerate}
\end{lem}
\proof
Let
$\mathcal{U}=\left\{ B(\alpha_{i},r)\mid i\in\mathbb{N},~r\in\mathbb{Q}^{+},~r<\varepsilon\right\}$.
Then $\mathcal{U}$ is an open cover of $(X,d)$ (since $\alpha $
is a dense sequence in $(X,d)$). The set $A$ is compact and therefore there exist $U_{1},\dots,U_{n}\in\mathcal{U}$
such that $A\subseteq U_{1}\cup\dots\cup U_{n}$.
We may assume that $U_{j}\cap A\not=\emptyset$
for each $j\in \{1,\dots,n\}$. Choose
$j_{1},\dots,j_{n}\in\mathbb{N}$ so that
$U_{k} =I_{j_{k} }$ and $\rho _{j_{k} }<\varepsilon $ for each $k\in \{1,\dots ,n\}$.
Let  $l\in\mathbb{N}$
be such that $[l]=\left\{ j_{1},\dots,j_{n}\right\} $. Then
$\langle A,l,\varepsilon \rangle $.

(ii)
Since $A$ and $B$ are compact, nonempty and disjoint, we have
$d(A,B)>0$.
Let $$\lambda = \min\left\{\varepsilon, \frac{d(A,B)}{4}\right\}.$$
Suppose  $j,j'\in \mathbb{N}$ and $A'\subseteq A$, $B'\subseteq B$ are such that
$\langle A',j,\lambda \rangle$ and $\langle B', j',\lambda \rangle$.
Let $i \in [j]$ and $i' \in [j']$. We claim that
\begin{equation}\label{eq-l1}
d(\lambda_i, \lambda_{i'}) > \rho_i + \rho_{i'}.
\end{equation}
Since $\langle A',j,\lambda \rangle$, we have $\rho_i < \lambda$ and
$I_i \cap A' \not= \emptyset$. Therefore there exists $a \in A$ such that
 $d(a,\lambda_i) < \rho_i$, hence  $d(a,\lambda_i) < \lambda $.
Similarly, $\rho _{i'}<\lambda $ and there exists $b \in B$ such that $
d(b, \lambda_{i'}) < \lambda $.

We have
$$ \rho _{i} +\rho _{i'}+2\lambda < 4 \lambda\leq d(A,B) \leq d(a,b) \leq d(a, \lambda_{i}) + d(\lambda_{i}, \lambda_{i'}) + d(\lambda_{i'}, b)
   <$$ $$< \rho_i + \rho_{i'} + d(\lambda_{i}, \lambda_{i'})<2\lambda + d(\lambda_{i}, \lambda_{i'})$$
Hence $\rho _{i} +\rho _{i'}+2\lambda<2\lambda + d(\lambda_{i}, \lambda_{i'})$ and (\ref{eq-l1}) follows.
The conclusion: $J_{j} $ and $J_{j'}$ are formally disjoint.
  \qed

\begin{lem}
\label{lem:effective-links} Let $(X,d,\alpha )$ be a computable metric space
and let $A_{1},\dots,A_{n}$ be compact nonempty sets in this space.
Let $\varepsilon>0$. Then there exist
 $j_{1},\dots,j_{n}\in \mathbb{N}$ such that
\begin{equation}\label{lem:effective-links-1}
 \langle A_{1},j_{1},\varepsilon\rangle,\mbox{ \dots, }\langle A_{n},j_{n},\varepsilon\rangle
\end{equation}
and
 such that for all $p,q\in\left\{ 1,\dots,n\right\} $ the following holds:
\[
(A_{p}\cap A_{q}=\emptyset\implies J_{j_p}\text{ and }J_{j_q}\text{ are formally disjoint}).
\]
\end{lem}
\proof
Let
\[
C=\left\{ (p,q)\in\left\{ 1,\dots,n\right\} \times\left\{ 1,\dots,n\right\} \mid A_{p}\cap A_{q}=\emptyset\right\} .
\]
 For each $(p,q)\in C$ by Lemma
 \ref{lem: lamba_td_Ji_Jj_formalno_disjunktni}
 there exists $\lambda_{(p,q)}>0$ such that
  $\lambda_{(p,q)} < \varepsilon$
and such that  $\langle A_{p},j,\lambda_{(p,q)}\rangle$ and $\langle A_{q},j',\lambda_{(p,q)}\rangle$
implies that $J_{j}$ and $J_{j'}$ are formally disjoint.
Let
$$\lambda=\min\left\{ \lambda_{(p,q)}\mid (p,q)\in C\right\}.$$
By Lemma \ref{lem: lamba_td_Ji_Jj_formalno_disjunktni} there exist $j_{1},\dots,j_{n}\in \mathbb{N}$
such that
\begin{equation}\label{lem:effective-links-2}
\langle A_{1},j_{1},\lambda\rangle,\mbox{ \dots, }\langle A_{n},j_{n},\lambda\rangle.
\end{equation}
If $p,q\in \{1,\dots ,n\}$ are such that $A_{p} \cap A_{q} =\emptyset $, then $(p,q)\in C$ and
$\lambda \leq \lambda _{(p,q)}$. Therefore $\langle A_{p},j_{p},\lambda\rangle$
and $\langle A_{q},j_{q},\lambda\rangle$ implies
$\langle A_{p},j_{p},\lambda_{(p,q)}\rangle$
and $\langle A_{q},j_{q},\lambda_{(p,q)}\rangle$ and this implies that
$J_{j} $ and $J_{j'}$ are formally disjoint. And (\ref{lem:effective-links-1}) clearly follows
from $\lambda <\varepsilon $ and (\ref{lem:effective-links-2}).  \qed

\begin{lem}
\label{lem:fdiam_ograda} Let $(X,d,\alpha)$ be a computable metric space.
Let $A\subseteq X$,  $j\in\mathbb{N}$ and $r>0$
be such that
 $\langle A,j,r\rangle$. Then $\mathrm{fdiam}(j)<4r+\mathrm{diam}\ A$.
 \end{lem}
\proof
Let $i,i',i''\in[j]$ be such that
\begin{equation}\label{lem:fdiam_ograda-1}
\fdiam(j)=d(\lambda_{i},\lambda_{i'})+2\rho_{i''}.
\end{equation}
Since $B(\lambda_{i},\rho_{i})\cap A\not=\emptyset$ there exists
$a\in A$ such that $d(\lambda_{i},a)<\rho_{i}$.
Also, there exists $b$ such that
$d(\lambda_{i'},b)<\rho_{i'}$.
Now $$d(\lambda_{i},\lambda_{i'})
\leq d(\lambda_{i},a)+d(a,b)+d(b,\lambda_{j})<\rho_{i}+\rho_{i'}+\mathrm{diam}\ A<2r+\mathrm{diam}\ A.$$
Using $\rho _{i''}<r$  and (\ref{lem:fdiam_ograda-1}) we get  $\mathrm{fdiam}(j)<4r+\mathrm{diam}\ A$. \qed

Let us now prove Proposition \ref{luk-formlanac}. Since $f$ is uniformly continuous, there
exists $n_{0} \in \mathbb{N}$, $n_{0} \geq 1$, such that $\diam(f([\frac{i}{n}r,\frac{i+1}{n}r]))<\frac{\varepsilon }{2}$
for all $i\in \{0,\dots ,n-1\}$ and $n\geq n_{0} $.

Fix $n\geq n_{0}$. Let $$A_{i}
=f\left(\left[\frac{i}{n}r,\frac{i+1}{n}r\right]\right)$$ for
$i\in \{0,\dots ,n-1\}$. By Lemma \ref{lem:effective-links} there
exist $j_{0} ,\dots ,j_{n-1}\in \mathbb{N}$ such that $\langle
A_{i} ,j_{i},\frac{\varepsilon }{8}\rangle $ for each $i\in
\{0,\dots ,n-1\}$ and such that $J_{j_{i} }$ and $J_{j_{i'}}$ are
formally disjoint for all $i,i'\in \{0,\dots ,n-1\}$ such that
$|i-i'|>1$. Finally, for each $i\in \{0,\dots ,n-1\}$ we have
$\diam A_{i} <\frac{\varepsilon }{2}$ and $\langle A_{i} ,j_{i}
,\frac{\varepsilon }{8}\rangle $ and it follows from Lemma
\ref{lem:fdiam_ograda} that $\fdiam(j_{i} )<\varepsilon $. \qed

\section{Co-c.e.\ topological rays} \label{sect-5}

A metric space $R$ is said to be a \textbf{topological ray} if $R$
is homeomorphic to $[0,\infty\rangle $. If $f:[0,\infty\rangle
\rightarrow R$ is a homeomorphism, then we say that $f(0)$ is an
\textbf{endpoint} of $R$.

In this section we prove that a semi-c.c.b.\  set $R$ must be
c.c.b.\  if $R$ is a topological ray with a computable endpoint.
Actually, we will prove a more general fact: if $R$ is a topological ray
with computable endpoint and if $R\cup F$ is semi-c.c.b, where $F$ is a closed set
disjoint with $R$, then $R$ is computably enumerable.

The first fact that we need here is that for such an $R$ the
following holds: if $f:[0,\infty\rangle \rightarrow R$ is a
homeomorphism, then $f(t)$ ``converges to infinity'' as $t$
converges to infinity. (In particular, $R$ is unbounded.) The
following proposition gives a precise description of this
property.

\begin{prop} \label{neom}
Let $(X,d)$ be a metric space. Let $R$ be a subset of $X$ such
that $R\cap B$ is a compact set for each closed ball $B$ in
$(X,d)$ and such that there exists a homeomorphism
$f:[0,\infty\rangle \rightarrow R$. Then for each closed ball $B$
there exists $t_{0} \in [0,\infty\rangle$ such that $f(t)\notin B$
for each $t\geq t_{0} $.
\end{prop}
\proof Suppose the opposite. Then there exists a closed ball
$B$ such that for each $t_{0} \in
[0,\infty\rangle$ there exists $t\geq t_{0} $ such that $f(t)\in B$.
Therefore there exists a sequence $(t_{n} )_{n\in \mathbb{N}}$ in $[0,\infty\rangle $
such that
$$t_{n} \geq n\mbox{ and }f(t_{n} )\in B$$
for each $n\in \mathbb{N}$.
Then clearly $f(t_{n} )\in R\cap B$ for each $n\in \mathbb{N}$ and since $R\cap B$ is compact,
there is a subsequence $(t_{n_{i} })_{i\in \mathbb{N}}$ of $(t_{n} )$ such
that the sequence $(f(t_{n_{i} }))$ converges to a point in $R\cap B$, hence it converges
to a point in $R$. However, since $f$ is homeomorphism (and $f^{-1} :R\rightarrow [0,\infty\rangle $ is continuous),
the sequence $(t_{n_{i} })$ converges to some point in $[0,\infty\rangle $,
which is impossible since this sequence is clearly unbounded ($i\leq n_{i}\leq t_{n_{i} }$ for each $i\in \mathbb{N}$). \qed

Note that the previous proposition does not hold without the
assumption that $R$ is compact on closed balls. For example, if $(X,d)$ is the real
line with the Euclidean metric and $R=[0,1\rangle $, then $R$ is
homeomorphic to $[0,\infty\rangle $, but $R$ is clearly bounded.

Suppose $R$ is a semi-c.c.b.\ topological ray with computable endpoint in some computable metric space. How to prove that $R$ is c.c.b., i.e.\ how to prove that $R$ is c.e.? Let $a$ be some fixed rational point which is close to the endpoint of $R$. We want, for given $n,k\in \mathbb{N}$, to effectively find finitely many rational open sets $C_{0} ,\dots ,C_{l} $ whose diameters are less then $2^{-k}$ and such that these sets cover $R\cap \widehat{B}(a,n)$ and  each of these sets intersects $R$. If we can do this, the fact that $R$ is c.e.\ will easily follow. Informally, we can imagine that the image of that part of $R$ which lies in  $\widehat{B}(a,n)$ becomes sharper and sharper as $k$ tends to infinity.

So how to get such sets $C_{0} ,\dots ,C_{l} $? Let
$f:[0,\infty\rangle \rightarrow R$ be a homeomorphism. By
Proposition \ref{neom} there exists $t_{0} >0$ such that $f(t)$
leaves $\widehat{B}(a,n)$ after $t=t_{0} $. (See Figure 1. The
blue curve is $f([0,t_{0} ])$. The black circle is the boundary of
$\widehat{B}(a,n)$.) \vspace{6pt}
\begin{center}
  \includegraphics[height=1.7in,width=2.3in]{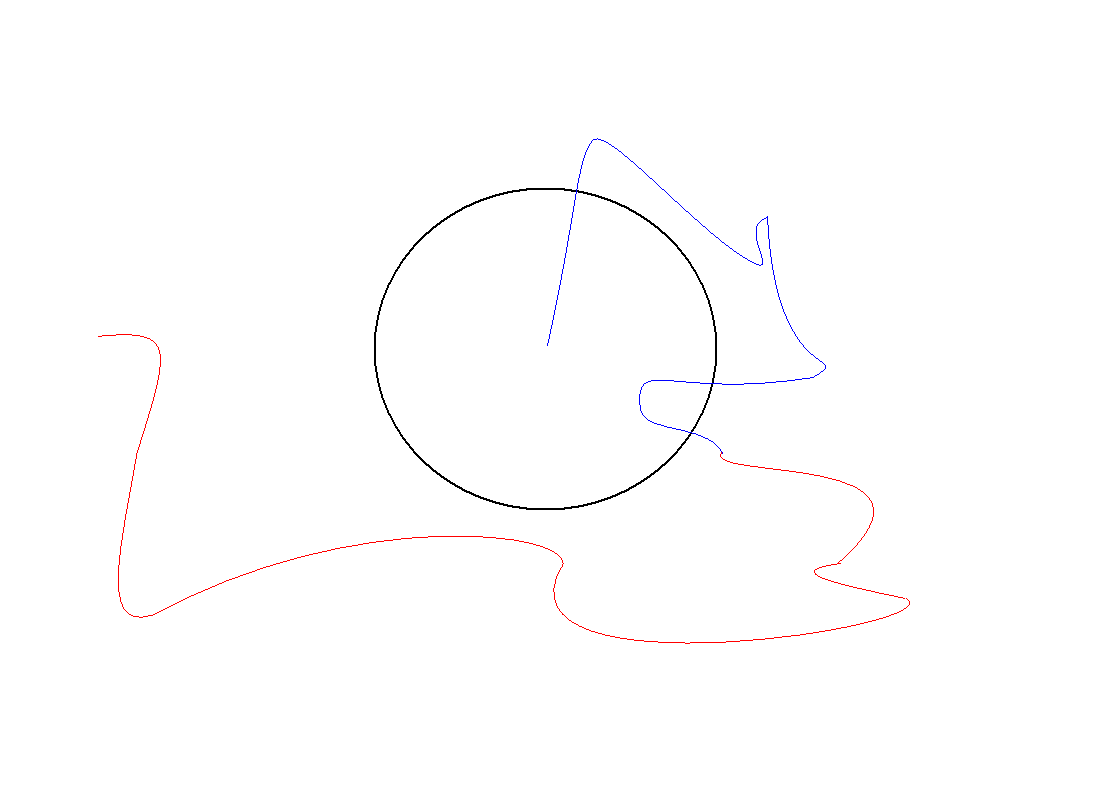}\hspace{35pt}
  \includegraphics[height=1.7in,width=2.3in]{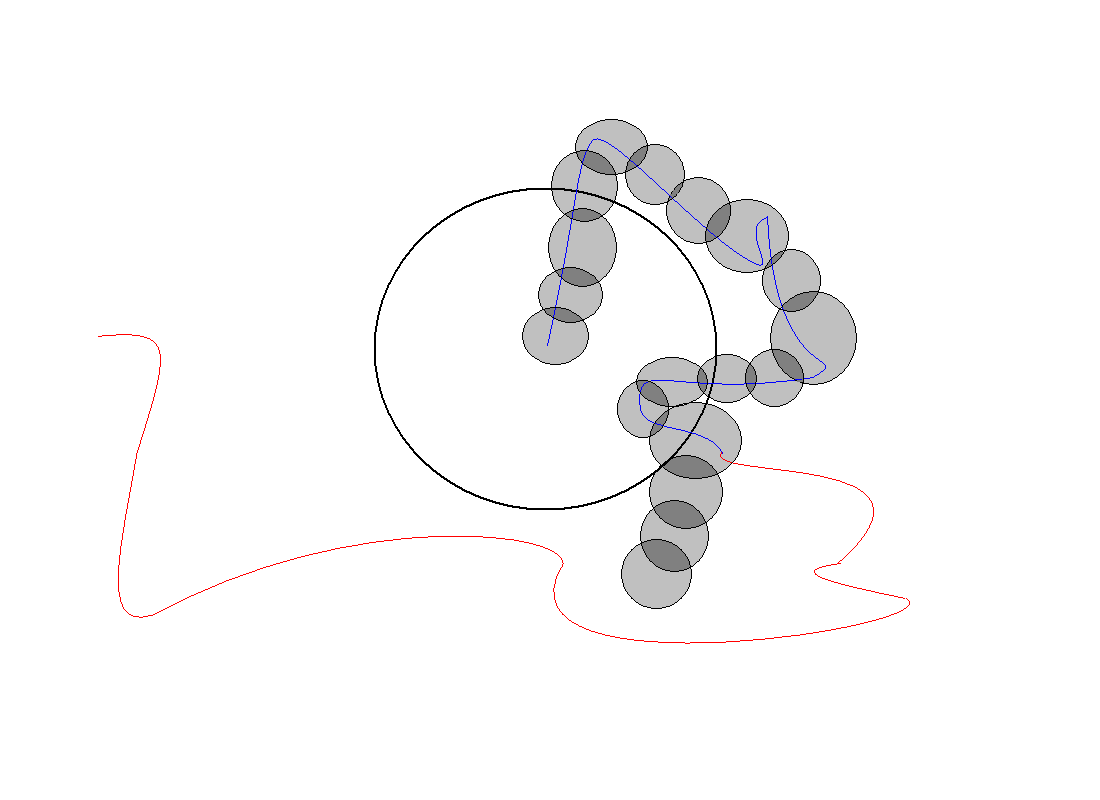}

  \emph{Figure 1.} \hspace{150pt} \emph{Figure 2.}
\end{center}
\vspace{6pt}
Therefore, $R\cap \widehat{B}(a,n)$ is contained in $f([0,t_{0} ])$ and this implies that there exists a rational $2^{-k}-$chain $C_{0} ,\dots,C_{l} $ which covers $R\cap \widehat{B}(a,n)$ and such that $f(0)\in C_{0} $.
(Figure 2.) These conditions are semi-decidable by results from Section \ref{sect-4} and therefore we can effectively find such a sequence of sets. However, we do not have the condition that each of these sets intersects $R$ and the question is does this follow from the conditions that we have?
The answer is no, as Figure 2 shows (the bottom three links do not intersect $R$).
So the question is what additional conditions to require on the chain $C_{0} ,\dots ,C_{l} $
so that these conditions are semi-decidable and so that they imply that each of the sets $C_{0} ,\dots ,C_{l} $
intersects $R$?

The idea is to proceed in the following way. Since $f([0,t_{0} ])$ is compact,
there exists $m\in \mathbb{N}$ such that $f([0,t_{0} ])\subseteq \widehat{B}(a,m)$. (See Figure 3. The green
circle is the boundary of $\widehat{B}(a,m)$.)
 Now
we can cover $R\cap \widehat{B}(a,m)$ (in the same way as we covered $R\cap \widehat{B}(a,n)$) by
a $2^{-k}-$chain $C_{0} ,\dots ,C_{p} $ such that $f(0)\in C_{0} $ (Figure 4). Again,
some of the sets $C_{0} ,\dots ,C_{l} $ may not intersect $R$ (the last three in Figure 4).
However, it will be possible to conclude
that for some $p\in \{0,\dots ,l-1\}$ the links $C_{0} ,\dots ,C_{p} $ cover $\widehat{B}(a,n)$
and they are all formally contained in $B(a,m)$ (these conditions are semi-decidable).
This altogether will imply that each of the links $C_{0} ,\dots ,C_{p} $ intersects $R$. (In Figure 4 $C_{0} ,\dots ,C_{m} $
are the links between blue links, including blue links.)
\vspace{6pt}
\begin{center}
  \includegraphics[height=1.7in,width=2.3in]{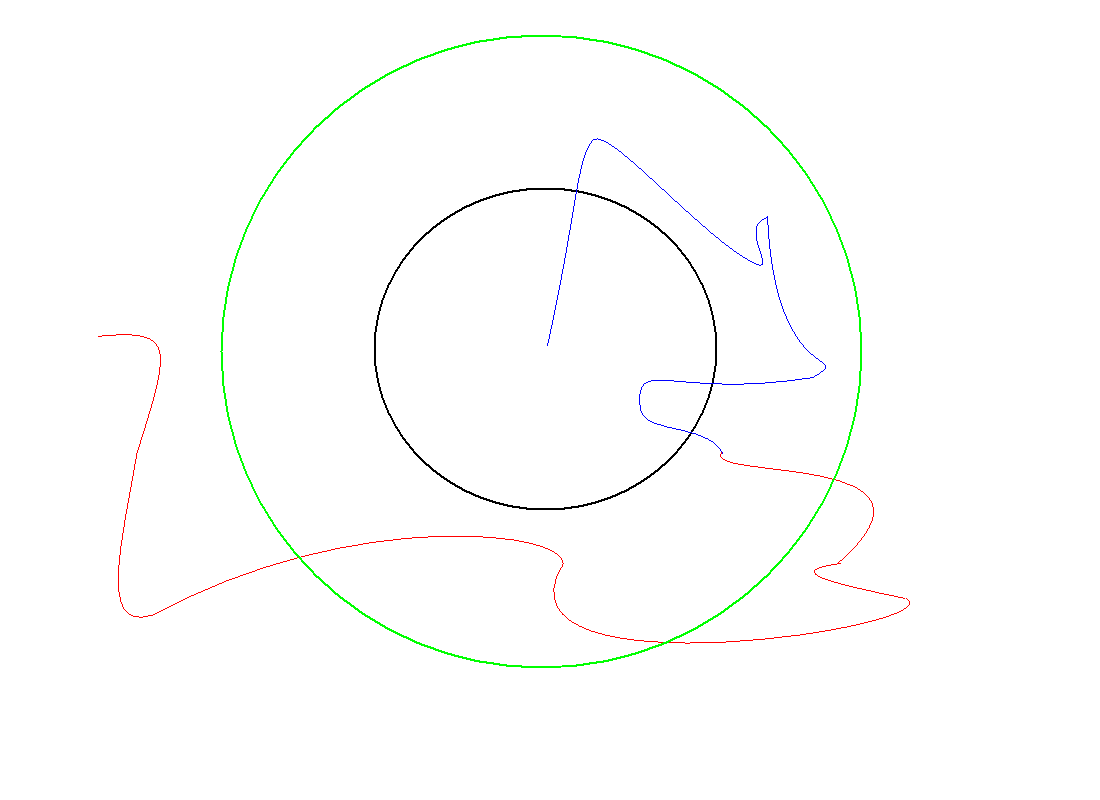}\hspace{35pt}
  \includegraphics[height=1.7in,width=2.3in]{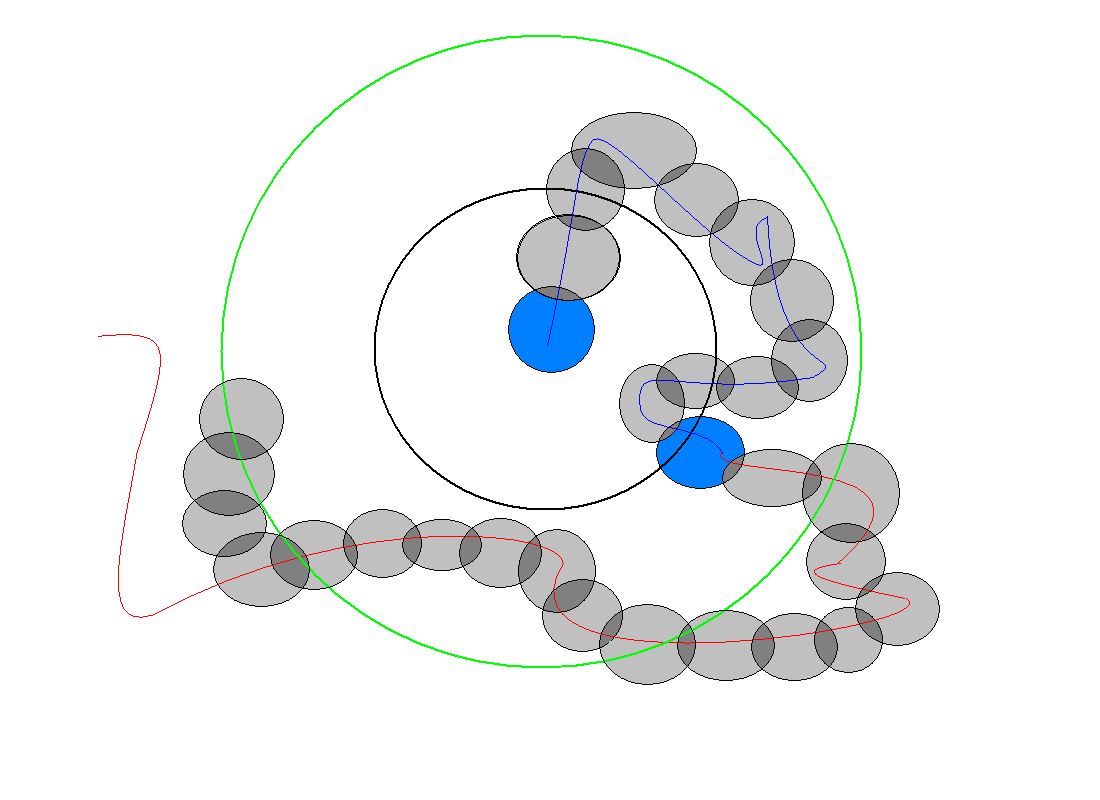}

  \emph{Figure 3.} \hspace{150pt} \emph{Figure 4.}
\end{center}
\vspace{6pt}

The described procedure is applied in the proof of the following theorem.

\begin{thm} \label{top-ray}
Let $(X,d,\alpha )$ be a computable metric space. Let $R$ be a
subset of $X$ which is, as a subspace of $(X,d)$, a
topological ray whose endpoint is computable. Suppose $F$ is a closed set in $(X,d)$ which is
disjoint with $R$ and such that $R\cup F$ is semi-c.c.b. Then $R$ is a computably enumerable (closed) set.
\end{thm}
\proof Let $f:[0,\infty\rangle \rightarrow R$ be a homeomorphism. Let $a$ be some
rational point such that $d(a,f(0))<1$.
For each $n\in \mathbb{N}$ let $$R_{n}=R\cap\widehat{B}(a,n)\mbox{ and }F_{n}=F\cap\widehat{B}(a,n).$$

Let $n,k\in\mathbb{N}$. By Proposition \ref{neom}
there exists $r>0$ such that $f(x)\not\in\widehat{B}(a,n)$ for each $x\geq r$.
The set $f([0,r])$ is compact and therefore there exists $m\in\mathbb{N}$, $m\geq 1$
such that
$f([0,r])\subseteq B(a,m)$.
Again, by Proposition \ref{neom} there exists
$r'>r$ such that $f(x)\not\in\widehat{B}(a,m)$, for each
$x\geq r'$.

Note that
\begin{equation}
  R_{n}\subseteq f([0,r])\subseteq R_{m}\subseteq f([0,r']).
  \label{eq:arc-inclusions}
\end{equation}
Let
\[
D=\max\left\{ d(a,f(x))\mid x\in[0,r]\right\}.
\]
Since
$f([0,r])\subseteq B(a,m)$, we have $D<m$.
Let
\[
\mu=\frac{m-D}{2}.
\]
By Lemma \ref{lem: lamba_td_Ji_Jj_formalno_disjunktni} there exists $\lambda >0$
such that if $j,j'\in \mathbb{N}$ and $A\subseteq f([0,r'])$, then
\begin{equation}\label{ray-11}
(\langle F_{m},j,\lambda \rangle \mbox{ and }\langle A,j',\lambda \rangle) \Longrightarrow
J_{j}\mbox{ and }J_{j'}\mbox{ are formally disjoint}.
\end{equation}

Let $\varepsilon =\min\{\mu ,\lambda\}$.
Let $u\in \mathbb{N}$ be such that $\langle F_{m} ,u,\lambda \rangle $.

Using Proposition \ref{luk-formlanac} we get $n'\geq 1$ and
  $j_0,\dots,j_{n'-1}\in \mathbb{N}$ so that
  \begin{enumerate}
   \item $\langle f([\frac{i}{n'}r',\frac{i+1}{n'}r']) ,j_{i},\varepsilon \rangle $ for each $i\in \{0,\dots ,n'-1\}$;
   \item $J_{j_i}$ and $J_{j_{i'}}$ are formally disjoint for all
          $i,i' \in \{0,\dots,n'-1\}$ such that $|i-i'|>1$;
    \item $\mathrm{fdiam}(j_i)<\min\{2^{-k},\varepsilon \}$ for each $i\in \{0,\dots,n'-1\}$.
  \end{enumerate}
It follows from (\ref{ray-11}) that $J_{u}$ and $J_{j_{i} }$ are
formally disjoint for each $i\in \{0,\dots ,n'-1\}$.

Let $\ell\in\mathbb{N}$ be such that
$$((\ell)_{0} ,\dots ,(\ell)_{\overline{\ell}})=(j_1,\dots,j_{(n'-1)}).$$
Then $\mathcal{H}_{\ell}$ is a formal chain which covers
$f([0,r'])$, $\fmesh(\ell)<2^{-k}$ and $J_{u}$ and
$\mathcal{H}_{\ell}$ are formally disjoint. Note that $f(0)\in
J_{(\ell)_{0} }$.

Since
$R_{m}\subseteq f([0,r'])$, we have
$R_{m}\subseteq\cup\mathcal{H}_{\ell}$.
Furthermore, since $r\in [0,r']$,
there exists $p\in\left\{ 0,\dots,\overline{\ell}\right\} $
such that $f(r)\in J_{(\ell)_{p}}$. The property (1) above
ensures
that
$f([0,r])\subseteq\cup\mathcal{H}_{\ell}^{0\leq p}$ and each
link of $\mathcal{H}_{\ell}^{0\leq p}$ intersects $f([0,r])$.
It follows
$R_{n}\subseteq\bigcup\mathcal{H}_{\ell}^{0\leq p}$.

We claim that  $\mathcal{H}_{\ell}^{0\leq p}$ is formally contained in
 $B(a,m)$. To see this, let us take $i\in \{0,\dots ,p\}$.
 We want to prove that
$J_{(\ell)_{i}}$ is formally contained in $B(a,m)$.

It would be enough to prove that
 $I_{k'}$ is formally contained in $B(a,m)$ for each $k'\in[(\ell)_{i}]$.
So let $k'\in[(\ell)_{i}]$. Since $J_{(\ell)_{i}}$ intersects
$f([0,r])$, there exists $b\in J_{(\ell)_{i}}$
such that $b\in f([0,r])$. Note that
$$d(\lambda _{k'},b)\leq \diam(J_{(\ell)_{i}})\leq \fdiam((\ell)_{i})\mbox{ and }\rho _{k'}\leq \fdiam((\ell)_{i}).$$
Also note that
\begin{equation}
  \mathrm{fdiam((\ell)_{i})}\leq\mathrm{fmesh}(\ell)
  <\varepsilon\leq \mu.\label{eq:fdiam_leq_mu}
\end{equation}
Therefore
$$d(\lambda _{k'},a)+\rho _{k'}\leq d(\lambda _{k'},b)+d(a,b)+\rho _{k'}\leq D+2\fdiam((\ell)_{i})<D+2\mu =m.$$
Hence $d(\lambda_{k'},a)+\rho_{k'}<m$ and this means that
$I_{k'}$ is formally contained in $B(a,m)$.

Finally, note that $p<\overline{\ell}$. Otherwise,  we would have
 $p=\overline{\ell}$. It is clear from the construction of the chain
 $\mathcal{H}_{\ell}$ (property (1)) that  $f(r')\in J_{(\ell)_{\overline{\ell}}}$.
 Hence $f(r')$ would belong to $J_{(\ell)_{p}}$.
However $f(r')\not\in\widehat{B}(a,m)$ which would contradict the fact
that $J_{(\ell)_{p}}$ is (formally) contained in $B(a,m)$.

We have the following conclusion.
For each $n,k\in\mathbb{N}$ there exist $\ell,m,p,u\in\mathbb{N}$
such that
\begin{enumerate}
\item \label{list:ray_H_ell_formal_chain}
       $\mathcal{H}_{\ell}$ is a formal chain;

\item \label{list:ray_H_ell_formal_chain-1}
       $J_{u}$ and $\mathcal{H}_{\ell}$ are formally disjoint;

\item \label{list:ray_a_in_J_ell_0}
       $f(0)\in J_{(\ell)_{0}}$;

\item \label{list:ray_Sn_subset_H_ell_leq_p}
       $R_{n}\cup F_{n} \subseteq\bigcup\mathcal{H}_{\ell}^{0\leq p}\cup J_{u};$

\item \label{list:ray_Sm_subseteq_H_ell}
       $R_{m}\cup F_{m} \subseteq\bigcup\mathcal{H}_{\ell}\cup J_{u};$

\item \label{list:ray_H_ell_leq_p_form_contained_B(a,m)}
       $\mathcal{H}_{\ell}^{0\leq p}$ is formally contained in $B(a,m)$;

\item \label{list:p_less_overline_ell}
       $p<\overline{\ell}$ and $m\geq 1$;

\item \label{list:ray_fmesh_ell_less_2kk0}
       $\mathrm{fmesh}(\ell)<2^{-k}$.
\end{enumerate}

Let
\[
  T=\left\{ (n,k,m,\ell,p,u)\in\mathbb{N}^{6}\mid \text{for }n,k,m,\ell,p,u
    \text{ properties (\ref{list:ray_H_ell_formal_chain})--(\ref{list:ray_fmesh_ell_less_2kk0}) hold}\right\} .
\]
Using Proposition \ref{formal-chain}, Lemma \ref{Jj-disj-Hl},
Proposition \ref{co-c.e.-semicomp-H-l} and Proposition \ref{NuR}
we conclude that $T$ is c.e.\ as the intersection of c.e.\ sets.
(Recall that $f(0)$ is computable point, and if $c$ is some
computable point, then it is straightforward to see that the set
$\{j\in \mathbb{N}\mid c\in J_{j} \}$ is c.e.)

We have shown that for all $n,k\in \mathbb{N}$
there exist $m,\ell,p,u\in\mathbb{N}$ such that
 $(n,k,m,\ell,p,u)\in T$.
 Therefore there exist  computable functions
$\widetilde{m},\widetilde{\ell},\widetilde{p},\widetilde{u}:\mathbb{N}^{2}\rightarrow\mathbb{N}$
such that $$(n,k,\widetilde{m}(n,k),\widetilde{\ell}(n,k),\widetilde{p}(n,k),\widetilde{u}(n,k))\in T$$
for all $n,k\in\mathbb{N}$ (Single-Valuedness Theorem).

Let $n,k\in \mathbb{N}$. Let $m=\widetilde{m}(n,k)$,
$\ell=\widetilde{\ell}(n,k)$, $p=\widetilde{p}(n,k)$ and $u=\widetilde{u}(n,k)$.
Then for $n,k,m,\ell,p,u$ properties (\ref{list:ray_H_ell_formal_chain})--(\ref{list:ray_fmesh_ell_less_2kk0}) hold.
Now we want to prove that each link of the chain
 $\mathcal{H}_{\ell}^{0\leq p}$ intersects $R$.

Notice first that there exists  $t\in[0,\infty\rangle$ such that
$d(a,f(t))=m$. Otherwise, $B(a,m)$ and $X\setminus\widehat{B}(a,m)$
would be disjoint open sets whose union contain $R$. However, each
of these sets intersects $R$, which follows from $d(a,f(0))<1$ and
Proposition \ref{neom} and so we would have that the topological
ray $R$ is disconnected, which is impossible.

The set $\{t\in[0,\infty\rangle\mid d(a,f(t))=m\}$ is a closed and
nonempty subset of $[0,\infty\rangle $ and therefore it has a minimal element.
Let $t_{0} $ be that element. Then $d(a,f(t))<m$ for each $t\in [0,t_{0} \rangle $
(if $f(t)>m$ for some $t\in [0,t_{0} \rangle $, then connectedness of
$f([0,t])$ implies that $d(a,f(s))=m$ for some $s\in [0,t]$ which is impossible
since $s<t_{0} $). Hence $f([0,t_{0} ])\subseteq R_{m} $.
It follows from property (\ref{list:ray_Sm_subseteq_H_ell})
that
$$f([0,t_{0} ])\subseteq \bigcup\mathcal{H}_{\ell}\cup J_{u}.$$
However $f([0,t_{0} ])\cap \bigcup\mathcal{H}_{\ell}\neq\emptyset $
by (\ref{list:ray_a_in_J_ell_0}) and $\bigcup\mathcal{H}_{\ell}\cap  J_{u}=\emptyset $
by (\ref{list:ray_H_ell_formal_chain-1}). The fact that $f([0,t_{0} ])$
is connected now gives
$$f([0,t_{0} ])\subseteq  \bigcup\mathcal{H}_{\ell}.$$

Therefore
$f(t_{0})\in J_{(\ell)_{v}}$ for some
$v\in\left\{ 0,\dots,\overline{\ell}\right\} $.
But now the property
(\ref{list:ray_H_ell_leq_p_form_contained_B(a,m)}) implies that $p<v$.
(If $v\leq p$, then
$J_{(l)_{v}}$ is (formally) contained in $B(a,m)$
 which is impossible since $f(t_{0})\in J_{(\ell)_{v}}$
 and $f(t_{0} )\notin B(a,m)$.)

Finally, let us prove that each link of the chain
 $\mathcal{H}_{\ell}^{0\leq p}$ intersects $R$.
 Suppose that there exists $i\in\left\{ 0,\dots,p\right\} $ such
 that
$J_{(\ell)_{i}}\cap R=\emptyset$.
Then  $i\not=0$ (since
$f(0)\in J_{(\ell)_{0}}$), hence $0<i<v$.
Now
$$U = J_{(\ell)_{0}}\cup\dots\cup J_{(\ell)_{i-1}}\mbox{ and }V = J_{(\ell)_{i+1}}\cup\dots\cup J_{(\ell)_{\overline{\ell}}}$$
are open disjoint sets  which cover $f([0,t_{0}])$
and each of these sets intersects $f([0,t_{0}])$
($f(0)\in J_{(\ell)_{0}}$, $f(t_{0})\in J_{(\ell)_{v}}$).
This is impossible since  $f([0,t_{0}])$
is connected.

So we have proved that $J_{(\ell)_{i}}\cap R\not=\emptyset$ for each
$i\in\left\{ 0,\dots,p\right\} $. Another fact regarding the chain
$\mathcal{H}_{\ell}^{0\leq p}$ that we want to verify is this:
if $s\in [0,\infty\rangle $ is such that $f([0,s])\subseteq B(a,n)$,
then $f(s)$ lies in some link of $\mathcal{H}_{\ell}^{0\leq p}$.

But if $s$ is such that $f([0,s])\subseteq B(a,n)$, then
$f([0,s])\subseteq R_{n} $ and now
(\ref{list:ray_Sn_subset_H_ell_leq_p}), together with the fact that
$f([0,s ])$ is connected, gives $f([0,s ])\subseteq \bigcup\mathcal{H}_{\ell}^{0\leq p}$.
In particular $f(s)$ lies in some link of $\mathcal{H}_{\ell}^{0\leq p}$.

We have the following conclusion: for all $n,k\in \mathbb{N}$
\begin{enumerate}
  \item the formal diameter of each link of the chain
$\mathcal{H}_{\widetilde{\ell}(n,k)}^{0\leq \widetilde{p}(n,k)}$ is less than $2^{-k}$;
  \item each link of the chain
$\mathcal{H}_{\widetilde{\ell}(n,k)}^{0\leq \widetilde{p}(n,k)}$ intersects $R$;
  \item if $s\in [0,\infty\rangle $ is such that $f([0,s])\subseteq B(a,n)$,
then $f(s)$ lies in some link of $\mathcal{H}_{\widetilde{\ell}(n,k)}^{0\leq \widetilde{p}(n,k)}$.
  \end{enumerate}

Note the following: if $c\in R$, then $c=f(s)$ for some $s\in [0,\infty\rangle $ and there exists $n\in \mathbb{N}$ such
that $f([0,s])\subseteq B(a,n)$. Then $f(s)$ (i.e.\ the point $c$) lies
in some link of $\mathcal{H}_{\widetilde{\ell}(n,k)}^{0\leq \widetilde{p}(n,k)}$
for each $k\in \mathbb{N}$.

Let $i\in \mathbb{N}$. Suppose $I_{i} \cap R\neq\emptyset $. Let $c\in I_{i} \cap R$.
Using Lemma \ref{fdiam-fdisj} we conclude that there exist $n,k\in \mathbb{N}$ such
that $c$ belongs to some link of $\mathcal{H}_{\widetilde{\ell}(n,k)}^{0\leq \widetilde{p}(n,k)}$
which is formally contained in $I_{i} $. So there exists $w\in \mathbb{N}$ such that
\begin{equation}\label{ray-12}
w\leq \widetilde{p}(n,k)\mbox{ and }J_{(\widetilde{\ell}(n,k))_{w}}\subseteq _{F}I_{i}.
\end{equation}
On the other hand, if (\ref{ray-12}) holds for some $n,k,w\in \mathbb{N}$, then $I_{i}$ intersects
$R$ because $J_{(\widetilde{\ell}(n,k))_{w}}$ does.
Hence $I_{i} \cap R\neq\emptyset $ if and only if there exist $n,k,w\in \mathbb{N}$ such that
(\ref{ray-12}) holds. It follows from Proposition \ref{fdiam-FD}(4) that $\{i\in \mathbb{N}\mid I_{i} \cap R\neq\emptyset \}$ is c.e.\
and this means that $R$ is c.e. \qed

\begin{cor} \label{top-ray-cor}
Let $(X,d,\alpha )$ be a computable metric space and let $R$ be a
semi-c.c.b.\  set in this space. Suppose $R$ is a topological ray  whose endpoint is computable. Then $R$ is c.c.b. \qed
\end{cor}

\section{Co-c.e.\ topological lines}\label{sect-6}

We will say that $L$ is a \textbf{topological line} if $L$ is a
metric space homeomorphic to $\mathbb{R}$.

While  we may imagine topological rays  as arcs which have one endpoint in infinity, a topological line can be thought of as an arc whose both endpoints are in infinity. And while for computability of a semi- c.c.b.\  topological ray we needed the assumption  that its endpoint is computable, in the case of a semi-c.c.b.\   topological line naturally  we will have no such  assumption. Hence we will prove  that each semi-c.c.b.\   topological line  is c.c.b. Actually, as in the case of topological rays, we will have a more general result.

 First,
we have a proposition similar to Proposition \ref{neom} which says that, under certain assumption, both tails of a topological line ``converge to infinity''.

\begin{prop} \label{neom-1}
Let $(X,d)$ be a metric space. Let $L$ be a subset of $X$ such
that $L\cap B$ is a compact set for each closed ball $B$ in
$(X,d)$ and such that there exists a homeomorphism $f:\mathbb{R}
\rightarrow L$. Then for each closed ball $B$ there exists $t_{0}
\in [0,\infty\rangle$ such that $f(t)\notin B$ for each $t\geq
t_{0} $ and $t\leq -t_{0} $.
\end{prop}
\proof The set $f([0,\infty\rangle )$ is closed in $L$. Therefore for each closed ball $B$ in $(X,d)$ the set $f([0,\infty\rangle )\cap B$ is closed in $L$ and consequently in $L\cap B$ which is compact. Hence $f([0,\infty\rangle )\cap B$  is compact. Similarly, $f(\langle -\infty,0] )\cap B$  is compact for each closed ball $B$ in $(X,d)$. Now we apply Proposition \ref{neom} on homeomorphisms  $[0,\infty\rangle\rightarrow f([0,\infty\rangle )$, $x\mapsto f(x)$, and  $[0,\infty\rangle\rightarrow f(\langle -\infty,0] )$, $x\mapsto f(-x)$. \qed

\begin{thm} \label{top-line}
Let $(X,d,\alpha )$ be a computable metric space. Let $L$ be a
subset of $X$ which is, as a subspace of $(X,d)$, a topological
line. Suppose $F$ is a closed set in $(X,d)$ which is disjoint
with $L$ and such that $L\cup F$ is semi-c.c.b. Then $L$ is a
computably enumerable set.
\end{thm}
\proof  Let $f:\mathbb{R} \rightarrow L$ be a homeomorphism. Let $a$ be some
rational point such that $d(a,f(0))<1$.
For each $n\in \mathbb{N}$ let $$L_{n}=L\cap\widehat{B}(a,n)\mbox{ and }F_{n}=F\cap\widehat{B}(a,n).$$

Let $\delta>0$ be such that
$f([-\delta, \delta]) \subseteq B(a,1)$. (Such a number
 exists since $f$ is continuous.)

Now choose  $A,B,C\in \mathbb{N}$ and $k_{0} \in \mathbb{N}$ so
that $f(-\delta )\in I_{A}$, $f(\delta )\in I_{B}$, $f(0)\in
I_{C}$, $\rho _{A}<\frac{2^{-k_0}}{4}$, $\rho
_{B}<\frac{2^{-k_0}}{4}$, $\rho _{C}<\frac{2^{-k_0}}{4}$ and
\begin{equation}\label{top-line-11}
2^{-k_{0}}<\min\left\{
d(I_{A},f([0,+\infty\rangle)),d(I_{B},f(\langle-\infty,0])),d(I_{C},F)\right\}.
\end{equation}

Let $n,k\in\mathbb{N}$. By Proposition \ref{neom-1}
there exists $r>0$ such that $f(x)\not\in\widehat{B}(a,n)$ for each $x\in \mathbb{R}$ such that
$x\geq r$ or $x\leq -r$.
Since $f([-r,r])$ is compact, there exists $m\in\mathbb{N}$, $m\geq 1$
such that
$f([-r,r])\subseteq B(a,m)$.
By Proposition \ref{neom-1} there also exists
$r'>r$ such that $f(x)\not\in\widehat{B}(a,m)$, whenever
$x\geq r'$ or $x\leq -r'$.

We have
\begin{equation}
  L_{n}\subseteq f([-r,r])\subseteq L_{m}\subseteq f([-r',r']).
  \label{eq:arc-inclusions-1}
\end{equation}
Let
\[
D=\max\left\{ d(a,f(x))\mid x\in[-r,r]\right\}.
\]
Then $D<m$ since $f([-r,r])\subseteq B(a,m)$.
Let
\[
\mu=\frac{m-D}{2}.
\]
By Lemma \ref{lem: lamba_td_Ji_Jj_formalno_disjunktni} there exists $\lambda >0$
such that if $j,j'\in \mathbb{N}$ and $G\subseteq f([-r',r'])$, then
\begin{equation}\label{line-111}
(\langle F_{m},j,\lambda \rangle \mbox{ and }\langle G,j',\lambda \rangle) \Longrightarrow
J_{j}\mbox{ and }J_{j'}\mbox{ are formally disjoint}.
\end{equation}

Let $\varepsilon =\min\{\mu ,\lambda, 2^{-(k + k_{0} + 3)}\}$.
Let $u\in \mathbb{N}$ be such that
\begin{equation}\label{line-1111}
\langle F_{m} ,u,\varepsilon  \rangle .
\end{equation}

Let $g : [0, 2r'] \rightarrow X$ be the function defined by
\[
   g(t) = f(t-r'),
\]
$t \in [0, 2r']$.

Applying Lemma \ref{luk-formlanac} to $g$, we get numbers
 $n'\geq 1$ and
  $j_0,\dots,j_{n'-1}\in \mathbb{N}$ such that
  \begin{enumerate}
    \item $\langle g([i \frac{2r'}{n'}, (i+1) \frac{2r'}{n'}]), j_i,\varepsilon \rangle $ for each $i\in
          \{0,\dots,n'-1\}$;
    \item \label{property:Ci_and_Cj_fdisj}
           $J_{j_i}$ are $J_{j_{i'}}$ formally disjoint for all
          $i,i' \in \{0,\dots,n'-1\}$ such that $|i-i'|>1$;
    \item \label{property:fdiam_less_than_epsilon}
    $\mathrm{fdiam}(j_i)<\varepsilon$ for each $i\in \{0,\dots,n'-1\}$.
  \end{enumerate}
\noindent
We can choose $n'$ so that $\frac{2r'}{n'} < \min\{\frac{r' - r}{2}, \frac{r}{2}\}$.
Let $\ell\in\mathbb{N}$ be such that
$$((\ell)_0, \dots, (\ell)_{\overline{\ell}}) = (j_0,\dots,j_{(n'-1)}).$$
Then  $\mathcal{H}_{\ell}$ is a formal chain and
$\fmesh(\ell) < \varepsilon $. It clearly covers $g([0,2r'])$, i.e.\ $f([-r',r'])$.
Hence $L_{m}\subseteq \bigcup \mathcal{H}_{\ell}$. And by
(\ref{line-111}) $J_{u}$ and $\mathcal{H}_{\ell}$ are formally disjoint.

Let $D'=\frac{2r'}{n'}$.
Let us choose numbers  $p,q,e \in \mathbb{N}$ so that
\begin{enumerate}
\item[(4)] $-r + r' \in [pD', (p+1)D']$;
\item[(5)] $r' \in [eD',
(e+1)D']$;
\item[(6)] $r+r'\in [qD', (q+1)D']$;
\end{enumerate}
Note that $f(-r)\in J_{j_{p} }$, $f(0)\in J_{j_{e} }$ and
$f(r)\in J_{j_{q} }$.

We claim that $p < e < q < \overline{\ell}$.
It holds $$pD' \leq r'-r \leq (e+1)D' - 2D' < eD'.$$
Dividing by  $D'$ we get $p < e$. Also
$$eD' \leq r' + r - r  \leq  (q+1)D' - r < (q+1)D' - 2D' < qD'$$
and we get $e<q$.

Let us prove that $q < \overline{\ell}$.
First we have
 $$(q+1) D' < qD' + \frac{r'-r}{2} \leq r+r' + \frac{r'-r}{2} = \frac{3r' + r}{2}  < 2r'.$$
Hence $(q+1) D' < 2r'$.
By definition of $\ell$ it holds $\overline{\ell} = n'- 1$.
Now
$$q D' = (q+1)D' - D' < 2r' - D' = (n' - 1) D' = \overline{\ell} D'$$
and it follows $q < \overline{\ell}$.

We claim that $I_{A}$ and $\mathcal{H}_{\ell}^{e\leq \overline{\ell}}$
are formally disjoint. Suppose the opposite.
Then there exists $i\in \{e,\dots , \overline{\ell}\}$ such that
 $I_{A}$ and $J_{(\ell)i}$ are not formally disjoint.
Therefore there exists  $j \in [(\ell)_i]$ such that
$$d(\lambda _{A}, \lambda_j) \leq \rho _{A} + \rho_j.$$
Note that by the construction of $\mathcal{H}_{\ell}$
each link of the chain $\mathcal{H}_{\ell}^{e\leq \overline{\ell}} $
intersects $f([0, \infty \rangle)$.
Therefore there exists $y\in J_{(\ell)_{i}} \cap f([0, \infty \rangle)$.
Now
\begin{align*}
  d(I_{A},f[0,\infty\rangle)  & \leq d(f(-\delta), y)  \\
                          & \leq d(f(-\delta), \lambda _{A} ) + d(\lambda _{A}, \lambda_j) + d(\lambda_j, y)\\
                          & \leq 2\rho _{A} + \rho_j + \diam (J_{(\ell)_{i}}) < 2\frac{2^{-k_{0} }}{4}+2\varepsilon < 2^{-k_0}
\end{align*}
which contradicts (\ref{top-line-11}).
Hence, $I_{A}$ and  $\mathcal{H}_{\ell}^{e\leq \overline{\ell}}$
are formally disjoint. In the same way we get
that $I_{B}$ and $\mathcal{H}_{\ell}^{0\leq  e}$
are formally disjoint and also, using (\ref{line-1111}), that $I_{C}$ and
$J_{u}$ are formally disjoint.

From the definition of numbers $p$ and $q$ we deduce that
$$ [-r, r]\subseteq  \bigcup_{p\leq i \leq q} [i D' - r', (i+1) D' - r']$$
which gives
$$f([-r,r])\subseteq  \bigcup_{p\leq i \leq q} f([i D' - r', (i+1) D' - r'])=\bigcup_{p\leq i \leq q} g([i D', (i+1) D'])\subseteq \bigcup \mathcal{H}_{\ell}^{p\leq q}.$$
Hence $$L_n \subseteq \bigcup \mathcal{H}_\ell^{p\leq q}.$$

Finally, let us prove that  $\mathcal{H}_\ell^{p \leq q}$
is formally contained in $B(a,m)$.

Let $i\in \{p,\dots ,q\}$.
 To prove that
$J_{(\ell)_{i}}$ is formally contained in $B(a,m)$ let us first
prove that $J_{(\ell)_{i}}$ intersects $f([-r,r])$.
Since $$g([i D', (i+1) D']) \subseteq J_{(\ell)_{i}}$$
it suffices to see that
$$[i D'-r', (i+1) D'-r']\cap [-r,r]\neq\emptyset .$$
For $i=p$ this intersection contains $-r$ and for $i=q$
it contains $r$. If $p<i$, then $p+1\leq i$ and $(p+1) D'-r'\leq i D'-r'$
which implies $-r\leq i D'-r'$. In the same way get that $i<q$ implies
$(i+1)D'-r'\leq r$. Hence if $i$ is between $p$ and $q$, then
the segment $[i D'-r', (i+1) D'-r']$ is contained in $[-r,r]$.

Now we proceed in the same way as in the proof of Theorem \ref{top-ray}.
We take $k'\in[(\ell)_{i}]$ and we want to prove that
 $I_{k'}$ is formally contained in $B(a,m)$.

Since $J_{(\ell)_{i}}$ intersects
$f([-r,r])$, there exists $b\in J_{(\ell)_{i}}$
such that $b\in f([-r,r])$. Then
$$d(\lambda _{k'},b)\leq \diam(J_{(\ell)_{i}})\leq \fdiam((\ell)_{i})\mbox{ and }\rho _{k'}\leq \fdiam((\ell)_{i}).$$
Also note that
\begin{equation}
  \mathrm{fdiam((\ell)_{i})}\leq\mathrm{fmesh}(\ell)
  <\varepsilon \leq \mu.\label{eq:fdiam_leq_mu}
\end{equation}
Therefore
$$d(\lambda _{k'},a)+\rho _{k'}\leq d(\lambda _{k'},b)+d(a,b)+\rho _{k'}\leq D+2\fdiam((\ell)_{i})<D+2\mu =m.$$
Hence $d(\lambda_{k'},a)+\rho_{k'}<m$ and
$I_{k'}$ is formally contained in $B(a,m)$.

The conclusion: for all $n,k\in\mathbb{N}$ there exist $m,\ell,p,q,e,u\in\mathbb{N}$
such that
\begin{enumerate}
  \item \label{list:H_ell_formal_chain}
  $\mathcal{H}_{\ell}$ is a formal chain;

  \item \label{list:H_ell_formal_chain-FDisj}
  $\mathcal{H}_{\ell}$ and $J_{u}$ are formally disjoint;

  \item \label{list:line_Sn_subset_H_ell_p_leq_q} $L_{n}\cup F_{n} \subseteq\cup\mathcal{H}_{\ell}^{p\leq q}\cup J_{u}$;

  \item \label{list:line_S_subset_Sm}
  $L_{m}\cup F_{m} \subseteq\bigcup\mathcal{H}_{\ell}\cup J_{u}$;

  \item \label{list:line_H_ell_form_contained_B(a,m)}
  $\mathcal{H}_{\ell}^{p\leq q}$ is formally contained in $B(a,m)$;

  \item \label{list:p_less_e_less_q_less_overline_ell}
  $p<e<q<\overline{\ell}$, $m\geq 1$;

  \item \label{list:fmesh_less_2kk0}
  $\mathrm{fmesh}(\ell)<2^{-(k+k_{0}+3)}$;

  \item \label{list:Jja_H_ell_geq_e_fdisj}
  $I_{A}$ and  $\mathcal{H}_{\ell}^{e\leq \overline{\ell} }$
  are formally disjoint;

  \item \label{list:line_Jjb_H_ell_leq_e_fdisj}
  $I_{B}$ and  $\mathcal{H}_{\ell}^{0\leq e }$
  are formally disjoint;

 \item \label{list:line_Jjb_H_ell_leq_e_fdisj-C}
  $I_{C}$ and  $J_{u}$
  are formally disjoint.
\end{enumerate}

Let $T$ be the set of all $(n,k,m,\ell,p,q,e,u)\in\mathbb{N}^{8}$
such that properties (\ref{list:H_ell_formal_chain})--(\ref{list:line_Jjb_H_ell_leq_e_fdisj-C})
hold. As in the proof of Theorem \ref{top-ray} we conclude
that $T$ is c.e.\ and we also conclude that there exists a computable function
$\varphi :\mathbb{N}^{2}\rightarrow \mathbb{N}^{6}$ such that
\begin{equation}\label{top-line-444}
(n,k,\varphi (n,k))\in T
\end{equation}
for all $n,k\in \mathbb{N}$.
This concludes the first part of the proof of Theorem \ref{top-line}.

In the second part we prove that the existence of a such function
$\varphi $ implies that $L$ is c.e.

Suppose we have $n,k,m,\ell,p,q,e,u\in\mathbb{N}$ such that
properties (\ref{list:H_ell_formal_chain})--(\ref{list:line_Jjb_H_ell_leq_e_fdisj-C}) hold. We also assume that $n\geq 1$.
We want to prove that each link of $\mathcal{H}_{\ell}^{p\leq q}$
intersects $L$. For $i\in \{0,\dots ,\overline{\ell}\}$ let $C_{i} =J_{(\ell)_{i} }$.
Hence $$\mathcal{H}_{\ell}=(C_{0} ,\dots ,C_{\overline{\ell}}).$$

First we prove the following: if  $t,s\in \mathbb{R}$ are such that
$t\leq 0\leq s$, then
\begin{enumerate}
 \item $f([t,s])\subseteq \widehat{B}(a,m)$ implies $f([t,s])\subseteq \bigcup \mathcal{H}_{\ell}$;

 \item $f([t,s])\subseteq \widehat{B}(a,n)$ implies $f([t,s])\subseteq \bigcup \mathcal{H}_{\ell}^{p\leq q}$.
 \end{enumerate}
If $f([t,s])\subseteq \widehat{B}(a,m)$, then $f([t,s])\subseteq L_{m} $, this and
(\ref{list:line_S_subset_Sm}) imply
$$f([t,s])\subseteq \bigcup\mathcal{H}_{\ell}\cup J_{u}$$
and $\bigcup\mathcal{H}_{\ell}$ and  $J_{u}$ are disjoint by
(\ref{list:H_ell_formal_chain-FDisj}). Since $f([t,s])$ is connected, it must
be entirely contained in one of these sets. But this cannot be $J_{u}$
since $f(0)\in f([t,s])$ and $f(0)$ belongs to $I_{C}$ which is
disjoint with $J_{u}$ by (\ref{list:line_Jjb_H_ell_leq_e_fdisj-C}).
Hence $f([t,s])\subseteq \bigcup \mathcal{H}_{\ell}$. In the same way we prove (2).

Since $f([-\delta ,\delta ])\subseteq B(a,1)\subseteq B(a,m)$,
there exist $\alpha ,\beta \in \{0,\dots ,\overline{\ell}\}$
such that $f(-\delta )\in C_{\alpha }$
and $f(\delta )\in C_{\beta  }$.

As in the proof of Theorem \ref{top-ray}, we conclude that there exist $s_{0} ,t_{0}\in \mathbb{R}$ such that
$s_{0} <0<t_{0} $, $d(a,f(s_{0} ))=d(a,f(t_{0} ))=m$ and $f(t)\in B(a,m)$
for each $t\in \langle s_{0} ,t_{0} \rangle $. It follows that there exist
$v,w\in \{0,\dots ,\overline{\ell}\}$ such that $f(s_{0} )\in J_{(\ell)_{v }}$,
$f(t_{0} )\in J_{(\ell)_{w }}$.

We claim that $p - 1\leq\alpha < e$ and $e<\beta\leq q+1$.

First, let us prove $p \leq \alpha + 1$. Suppose the opposite. Then
$\alpha + 1 < p < q$.
The link $C_\alpha$ is then disjoint with each of the links $C_p$, \dots, $C_q$.
However $f([-\delta ,\delta ])\subseteq B(a,n)$ since $n\geq 1$,
therefore $f(-\delta )\in C_{p}\cup \dots \cup C_{q}$ and, by definition of $\alpha $,
$f(-\delta )\in C_{\alpha }$. A contradiction. Hence, $p \leq \alpha + 1$.

Let us prove $\alpha < e$. Suppose the opposite. Then $\alpha \geq e$,
hence the link $C_\alpha$ is one of the links $C_e,\dots,C_{\overline{\ell}}$
and $f(-\delta) \in C_\alpha$.
On the other hand, $f(-\delta) \in I_{A}$ and this now
contradicts (\ref{list:Jja_H_ell_geq_e_fdisj}).
So $\alpha < e$ and altogether
$$p - 1 \leq \alpha < e.$$

In the same way we get
$$e < \beta \leq q+1.$$

Now we claim that $v<p$ and $q<w$. Let us prove $v<p$.

Suppose $p\leq v$.
This implies $q < v$. Otherwise we have $v\leq q$, which together with
$p\leq v$ means that $C_{v}$ is one of the links of the chain
 $\mathcal{H}_\ell^{p\leq q}$. But this chain
is formally contained in $B(a,m)$, hence $C_{v}\subseteq B(a,m)$.
This is impossible since   $f(s_{0} )\in C_{v}$.

Hence $p<q<v$. So $q+1\leq v$ which together with
$\beta \leq q + 1$ gives
$\beta\leq v$.
But $\beta\not=v$ because $\beta = v$ would imply
\[
 d(f(s_0), f(\delta)) < \mathrm{diam}\ C_v < 2^{-k_0},
\]
and this is impossible by (\ref{top-line-11}).
Therefore $\beta<v$.

We also have
\begin{equation}\label{top-line-222}
f([s_{0},-\delta])\cap C_{\beta}=\emptyset.
\end{equation}
Otherwise, there exists $y\in f([s_{0},-\delta])\cap C_{\beta}$ and
$$ d(I_{B},f(\langle-\infty,0]))\leq d(f(\delta),f(\langle-\infty,0]))\leq d(f(\delta),f([s_{0},-\delta]))\leq $$
$$\leq d(f(\delta),y)\leq\mathrm{diam}\ C_{\beta}<2^{-k_{0}}$$
which again contradicts (\ref{top-line-11}). Hence (\ref{top-line-222}) holds.

Let $U$ and $V$ be defined by
\[
 U= \bigcup_{0 \leq i\leq \beta -1} C_i, \quad
 V = \bigcup_{\beta+1 \leq i \leq \overline{\ell}} C_i.
\]
Since $\mathcal{H}_\ell$  covers $f([s_{0},t_0])$
and (\ref{top-line-222}) holds,
\begin{equation}\label{top-line-33}
f([s_{0}, -\delta ]) \subseteq U \cup V.
\end{equation}
We have $f(-\delta)\in C_\alpha$ and $\alpha < e < \beta$,
hence
\begin{equation}\label{top-line-331}
f([s_{0}, -\delta ])\cap U\neq\emptyset.
\end{equation}
 Furthermore,
$f(s_0)\in C_{v}$ and $\beta < v$, so
\begin{equation}\label{top-line-332}
f([s_{0}, -\delta ])\cap V\neq\emptyset.
\end{equation}
Finally, (\ref{list:H_ell_formal_chain}) implies that
 $C_i\cap C_{i'}=\emptyset $
whenever $i,i'\in \{0,\dots ,\overline{\ell}\}$ are such that $i<\beta <i'$.
Hence
\begin{equation}\label{top-line-333}
U\cap  V=\emptyset.
\end{equation}
From (\ref{top-line-33}), (\ref{top-line-331}), (\ref{top-line-332})
and (\ref{top-line-333}) it follows that $f([s_{0}, -\delta ])$ is not connected. A contradiction.

So we have proved that $v<p$. In the same way we get $q < w$.
Hence
\[
  v<p < q<w.
\]
It is easy to conclude from this that each link of the chain
 $\mathcal{H}_{\ell}^{p\leq q}$ intersects $L$.
Namely, let $i\in\mathbb{N}$ be such that $p\leq i\leq q$.
Then $v<i<w$.
Suppose that
$C_{i}\cap  L=\emptyset $. Then
$$U=C_{0}\cup\dots\cup C_{i-1}\mbox{ and } V=C_{i+1}\cup\dots\cup C_{\overline{\ell}}$$
are disjoint sets, their union covers $f([s_{0},t_{0}])$ and each
of these sets intersects $f([s_{0},t_{0}])$ because $f(s_{0} )\in C_{v}\subseteq U$ and
$f(t_{0} )\in C_{w}\subseteq V$. This contradicts the fact that
$f([s_{0},t_{0}])$ is connected.

Hence each link of the chain $\mathcal{H}_{\ell}^{p\leq q}$
intersects  $L$ (under the assumption that $n\geq 1$).

Let $\widetilde{m},\widetilde{\ell},\widetilde{p},\widetilde{q},\widetilde{e},\widetilde{u}:\mathbb{N}^{2}\rightarrow \mathbb{N}$
be the component functions of the function $\varphi $ from (\ref{top-line-444}).

If $c\in L$, then $c\in f([-t,t])$ for some $t\geq 0$.
Choose $n\in \mathbb{N}$, $n\geq 1$, so that
$f([-t,t])\subseteq B(a,n)$. Then for each $k\in \mathbb{N}$
some link of the chain $\mathcal{H}_{\widetilde{\ell}(n,k)}^{\widetilde{p}(n,k)\leq \widetilde{q}(n,k)}$
contains $c$.

Let $i\in \mathbb{N}$. As in the proof of Theorem \ref{top-ray} we
conclude that $I_{i} \cap L\neq\emptyset $ if and only if there
exist $n,k,w\in \mathbb{N}$ such that
$$\widetilde{p}(n,k)\leq
w\leq \widetilde{q}(n,k),~n\geq 1\mbox{ and
}J_{(\widetilde{\ell}(n,k))_{w}}\subseteq _{F}I_{i}.$$
Therefore
$L$ is c.e. \qed

\begin{cor} \label{top-line-cor}
Let $(X,d,\alpha )$ be a computable metric space and let $L$ be a
semi-c.c.b.\  set in this space. Suppose $L$ is a topological line.
Then $L$ is c.c.b. \qed
\end{cor}

\section{1-manifolds} \label{sect-7}
A 1-\textbf{manifold with boundary} is a second countable
Hausdorff topological space $X$ in which each point has a
neighborhood homeomorphic to $[0,\infty\rangle $. The
\textbf{boundary} $\partial  X$ of $X$ consists of those points
$x\in X$ for which every homeomorphism between a neighborhood of
$x$ and $[0,\infty\rangle $ maps $x$ to $0$. Therefore, each point
of $X\setminus \partial  X$ has a neighborhood in $X$ which is
homeomorphic to $\mathbb{R}$. If $\partial  X=\emptyset $, then we
simply say that $X$ is a 1-\textbf{manifold}.

If $X$ and $Y$ are topological spaces and $f:X\rightarrow Y$ a homeomorphism and
if $X$ is a 1-manifold with boundary, then $Y$ is also and $\partial Y=f(\partial X)$.

For example, $\mathbb{R}$ and the unit circle $S^{1}$ in $\mathbb{R}^{2}$ are
 1-manifolds, while $[0,\infty\rangle $ and $[0,1]$ are 1-manifolds with boundary,
 $\partial [0,\infty\rangle =\{0\}$, $\partial [0,1]=\{0,1\}$.
Each topological line is a 1-manifold and if $R$ is a topological
ray and $a$ is its endpoint, then $R$ is a 1-manifold with boundary and $\partial R=\{a\}$.
Furthermore, if $S$ is an arc with endpoints $a$ and $b$, then $S$
is a manifold with boundary and $\partial S=\{a,b\}$.
Note the following: if a subspace $M$ of some topological space $X$ is a
manifold with boundary, then
the boundary of $M$ in general differs from the
topological boundary of $M$ in $X$.

Since $a$ is a computable point if and only if $\{a\}$ is c.c.b.,
Theorem \ref{top-ray} means that a semi-c.c.b.\ topological
ray is c.c.b.\ if its boundary is c.c.b. The natural question arises whether
this holds  for each 1-manifold, i.e.\ if $M$ is a semi-c.c.b.\
1-manifold in a computable metric space, does the implication
\begin{equation}\label{mnf-impl}
\partial M\mbox{ c.c.b.}\Rightarrow M\mbox{ c.c.b.}
\end{equation}
hold? The answer is no, implication (\ref{mnf-impl}) fails to be
true in general.

 To see this, let $S$ be a c.e.\ subset of $\mathbb{N}$ which is
 not computable. The fact that $S$ is c.e.\ implies that the set
 $T=\mathbb{N}\setminus S$ is co-c.e.\ in $\mathbb{R}$. Therefore
 $T\times \mathbb{R}$ is co-c.e.\ in $\mathbb{R}^{2}$. Let $M=T\times
 \mathbb{R}$. Since $T
 \subseteq \mathbb{N}$, we have that $M$ is a
 1-manifold. That $M$ is not computable in $\mathbb{R}^{2}$ can be deduced
 from the fact that $T$ is not computable in $\mathbb{N}$. Of course $M$ is
 semi-c.c.b.\ by Proposition \ref{prop-co-c.e.-semi} and we
 conclude that (\ref{mnf-impl}) does not hold (note that $\partial
 M=\emptyset $).

However, we will show later that (\ref{mnf-impl}) holds under
additional assumption that $M$ has finitely many components.

It is known (see e.g.\ \cite{sh}) that if $X$ is a connected
1-manifold with boundary, then $X$ is homeomorphic to
$\mathbb{R}$, $[0,\infty\rangle $, $[0,1]$ or $S^{1}$. (Here
$S^{1}$ denotes the unit circle in $\mathbb{R}^{2}$.) Hence
topological lines, topological rays, arcs and topological circles
are all connected 1-manifolds.

 It is easy to conclude that if $X$
is a 1-manifold with boundary, then each component of $X$ is also
a 1-manifold with boundary and $x\in X$ belongs to the boundary of
$X$ if and only if $x$ belongs to the boundary of some component
of $X$.

\begin{thm} \label{komponenta}
Let $(X,d,\alpha )$ be a computable metric space. Suppose $M$ is a
semi-c.c.b.\ set which is a 1-manifold with boundary. Let $K$ be a
component of $M$.
\begin{enumerate}
\item If $K$ is a topological line or a topological circle, then
$K$ is c.e.\

\item If $K$ is a topological ray with computable endpoint or an
arc with computable endpoints, then $K$ is c.e.
\end{enumerate}
\end{thm}
\proof Let $x\in K$. Then $x$ has a neighborhood in $M$ which is
homeomorphic to $[0,\infty\rangle $. Hence $x$ has a neighborhood
in $M$ which is connected and which therefore is contained in $K$.
This means that $x$ belongs to some set which is open in $M$ and
is contained in $K$. So the conclusion is that $K$ is open in $M$.

Let $F=M\setminus K$. Then $F$ is closed in $M$, but since $M$ as
a semi-c.c.b.\ set is closed in $(X,d)$, we have that $F$ is
closed in $(X,d)$. Hence $F$ is closed, disjoint with $K$  and
$F\cup K$ is semi-c.c.b. Now Theorem \ref{top-ray} and Theorem
\ref{top-line} imply that $K$ is c.e.\ if $K$ is a topological ray
with computable endpoint or a topological line.

Suppose now that $K$ is a topological circle or an arc with
computable endpoints. Then $K$ is compact and since it is disjoint
with $F$ (which is closed), there exist $i_{0} ,\dots ,i_{n} \in
\mathbb{N}$ such that $$K\subseteq \widehat{I}_{i_{0} }\cup \dots
\cup \widehat{I}_{i_{n} }\subseteq X\setminus F.$$ Then we have
$$K=K\cap (\widehat{I}_{i_{0} }\cup \dots
\cup \widehat{I}_{i_{n} })=(K\cup F)\cap (\widehat{I}_{i_{0} }\cup \dots
\cup \widehat{I}_{i_{n} })=(M\cap \widehat{I}_{i_{0} })\cup \dots \cup (M\cap \widehat{I}_{i_{n} }) .$$
So for $j\in
\mathbb{N}$ the following equivalence holds:
$$K\subseteq J_{j}\Leftrightarrow M\cap \widehat{I}_{i_{0} }\subseteq J_{j} ,\dots ,M\cap \widehat{I}_{i_{n} }\subseteq J_{j}.$$
From this and the fact that $M$ is semi-c.c.b.\ we conclude that $K$ is semi-computable
compact set. Hence $K$ is a compact manifold with computable boundary and therefore,
by \cite{lmcs:mnf}, $K$ is a computable compact set. In particular, $K$ is c.e.\ \qed

As we have seen, if $M$ is a 1-manifold with boundary such that $M$ is semi-c.c.b.\
and $\partial M$ is c.c.b., then $M$ need not be c.c.b. Since $M$ is already semi-c.c.b.,
this means that $M$  need not be
computably enumerable. However,  although $M$ is not necessarily computably enumerable,
each component of $M$ is computably enumerable.

\begin{thm} \label{komp-ce}
Let $(X,d,\alpha )$ be a computable metric space. Let $M$ be
a 1-manifold with boundary in this space and suppose $M$ and $\partial M$ are
semi-c.c.b. Then each component of $M$ is computable enumerable.
\end{thm}
\proof In view of Theorem \ref{komponenta} it suffices to prove that each
point in $\partial M$ is computable. Let $x\in \partial M$. Then $x$ has a
neighborhood $N$ in $M$ such that there exists a
homeomorphism  $f:N\rightarrow [0,\infty\rangle $ such that $f(x)=0$. It is clear
from this that $x$ is the only point in $N$ which belongs to the boundary of $M$.
It follows that $B(x,r)\cap \partial M=\{x\}$ for some $r>0$ and we conclude
from this that $\widehat{I}_{i}\cap \partial M=\{x\}$ for some $i\in \mathbb{N}$.
Since $\partial M$ is semi-c.c.b., $\widehat{I}_{i}\cap \partial M$ is clearly
semi-computable compact set, hence $\{x\}$ is semi-computable and consequently $x$ is a computable
point. \qed

Since the union of finitely many c.e.\ sets in $(X,d,\alpha )$ is a c.e.\ set, we have the following
theorem.

\begin{thm} \label{glavni}
Let $(X,d,\alpha )$ be a computable metric space. Let $M$ be a subset of $X$ which is,
as a subspace of $(X,d)$,
a 1-manifold with boundary which has finitely many components.  Suppose $M$ and $\partial M$ are
semi-c.c.b. Then $M$ is c.c.b. \qed
\end{thm}

\begin{cor}
Let $(X,d,\alpha )$ be a computable metric space. Let $M$ be a 1-manifold
in this space and suppose $M$ has finitely many components and  $M$ is
semi-c.c.b. Then $M$ is c.c.b. \qed
\end{cor}

The following theorem is an immediate consequence of Theorem \ref{glavni}
and Proposition \ref{prop-co-c.e.-semi}.
\begin{thm} \label{gl-1}
Let $(X,d,\alpha )$ be a computable metric space which has compact closed balls
and the effective covering property. Let $M$ be a 1-manifold with boundary
in this space such that $M$ has finitely many components.  Suppose $M$ and $\partial M$ are
co-c.e. Then $M$ is computable. \qed
\end{thm}

\begin{cor} \label{gl-2}
Let $(X,d,\alpha )$ be a computable metric space which has compact closed balls
and the effective covering property. Let $M$ be a 1-manifold
in this space and suppose $M$ has finitely many components and  $M$ is
co-c.e. Then $M$ is computable.  \qed
\end{cor}

Finally, let us mention that Theorem \ref{gl-1} and Corollary \ref{gl-2}
do not hold in a general computable metric space. In \cite{glasnik}
an example of a computable metric space $(X,d,\alpha )$ can be found in which there exist a
co-c.e.\ arc with computable endpoints which is not computable and a co-c.e.\ topological circle which is
not computable. Moreover, we can find such $(X,d,\alpha )$ so that $(X,d,\alpha )$ has compact closed balls
and we can also find such $(X,d,\alpha )$ so that $(X,d,\alpha )$ has the effective covering property (but of course not
with both of these properties at the same time).

\section*{Acknowledgements}
The authors are grateful to anonymous referees for their useful suggestions
and corrections.

\end{document}